\documentclass[sigconf]{acmart}

\newcommand{\myparatight}[1]{\smallskip\noindent{\bf {#1}:}~}

\usepackage{pgf}
\usepackage{xcolor}
\usepackage{tikz}
\usetikzlibrary{tikzmark}

\usepackage{float}
\usepackage{stfloats}

\usepackage{newfloat}
\usepackage{listings}

\usepackage{bibentry}

\usepackage[skip=0pt]{caption}

\usepackage{amsmath,amsfonts}
\usepackage{xcolor}
\usepackage{multirow}
\usepackage{array} 
\usepackage{textcomp} 
\usepackage{float}
\usepackage{amsthm}

\usepackage{url}
\usepackage{verbatim}
\usepackage{graphicx} 
\usepackage{subcaption}
\usepackage{makecell}
\usepackage{bm}
 
\usepackage{algorithm}
\usepackage{algpseudocode}

\usepackage{xspace}

\usepackage{color, colortbl}

\definecolor{greyL}{RGB}{230,248,255}

\newcommand{\alg}{\texttt{SecureSplit}\xspace}

\usepackage{balance}

\copyrightyear{2026}
\acmYear{2026}
\setcopyright{cc}
\setcctype{by}
\acmConference[WWW '26]{Proceedings of the ACM Web Conference 2026}{April 13--17, 2026}{Dubai, United Arab Emirates}
\acmBooktitle{Proceedings of the ACM Web Conference 2026 (WWW '26), April 13--17, 2026, Dubai, United Arab Emirates}
\acmPrice{}
\acmDOI{10.1145/3774904.3792484}
\acmISBN{979-8-4007-2307-0/2026/04}

\begin{document}

\title{\alg: Mitigating Backdoor Attacks in Split Learning}

\author{Zhihao Dou}
\authornote{Equal contribution. Zhihao Dou conducted this research while he was an intern under the supervision of Minghong Fang.}
\affiliation{
	\institution{Case Western Reserve University}
	\city{Cleveland}
	\country{USA}
}

\author{Dongfei Cui}
\authornotemark[1]
\affiliation{
	\institution{Northeast Electric Power University}
	\city{Jilin}
	\country{China}
}

\author{Weida Wang}
\authornotemark[1]
\affiliation{
	\institution{Fudan University}
	\city{Shanghai}
	\country{China}
}

\author{Anjun Gao}
\affiliation{
	\institution{University of Louisville}
	\city{Louisville}
	\country{USA}
}

\author{Yueyang Quan}
\affiliation{
	\institution{University of North Texas}
	\city{Denton}
	\country{USA}
}

\author{Mengyao Ma}
\affiliation{
	\institution{The University of Queensland}
	\city{Brisbane}
	\country{Australia}
}

\author{Viet Vo}
\affiliation{
	\institution{Swinburne University of Technology}
	\city{Melbourne}
	\country{Australia}
}

\author{Guangdong Bai}
\affiliation{
	\institution{City University of Hong Kong}
	\city{Hong Kong}
	\country{China}
}

\author{Zhuqing Liu}
\affiliation{
	\institution{University of North Texas}
	\city{Denton}
	\country{USA}
}

\author{Minghong Fang}
\affiliation{
	\institution{University of Louisville}
	\city{Louisville}
	\country{USA}
}

\renewcommand{\shortauthors}{Zhihao Dou et al.}

\begin{abstract}
Split Learning (SL) offers a framework for collaborative model training that respects data privacy by allowing participants to share the same dataset while maintaining distinct feature sets. However, SL is susceptible to backdoor attacks, in which malicious clients subtly alter their embeddings to insert hidden triggers that compromise the final trained model. To address this vulnerability, we introduce \alg, a defense mechanism tailored to SL. \alg applies a dimensionality transformation strategy to accentuate subtle differences between benign and poisoned embeddings, facilitating their separation. With this enhanced distinction, we develop an adaptive filtering approach that uses a majority-based voting scheme to remove contaminated embeddings while preserving clean ones. Rigorous experiments across four datasets (CIFAR-10, MNIST, CINIC-10, and ImageNette), five backdoor attack scenarios, and seven alternative defenses confirm the effectiveness of \alg under various challenging conditions.
\end{abstract}



\maketitle


\section{Introduction} 
\label{sec:intro}

As data privacy concerns grow, federated learning (FL)~\cite{mcmahan2017communication} has gained attention for enabling collaborative model training without exposing raw data. FL allows multiple clients to train a shared model while keeping their data local, and is typically categorized into horizontal federated learning (HFL)~\cite{webank,paulik2021federated,chen2024contributions,xie2022federatedscope} and split learning (SL)~\cite{fu2022blindfl,romanini2021pyvertical,liu2020asymmetrical,thapa2022splitfed,poirot2019split,vepakomma2018split,singh2019detailed}. 
HFL applies when clients share the same feature space but hold different samples; for example, banks collaboratively train a credit scoring model using similar features but from distinct user groups.
In contrast, SL applies when each client holds only a subset of the features, while the server owns the labels. For example, one institution may store income data while another holds repayment records. Each client trains a {\em bottom model} to convert local features into embeddings, which are sent to the server. The server uses these embeddings and labels to update a {\em top model}, and then returns gradients to refine the clients' bottom models via backpropagation.

Although SL is decentralized, it remains vulnerable to poisoning attacks, particularly backdoor attacks~\cite{bai2023villain,he2023backdoor,naseri2024badvfl}. In such attacks, adversaries control malicious clients that generate embeddings crafted to manipulate the top model. These clients embed backdoor triggers into their local data or transmitted embeddings, causing the top model to misclassify specific inputs into attacker-chosen labels. For instance, the VILLAIN attack~\cite{bai2023villain} leverages label inference to locate target-class samples and subtly inserts a stealthy trigger into the embeddings of compromised clients.
A common countermeasure is to adopt Byzantine-robust aggregation~\cite{yin2018byzantine,blanchard2017machine,fang2024byzantine,fang2025we,cao2020fltrust,fang2025provably} from HFL to mitigate backdoor attacks. However, our experiments show that directly applying these defenses in SL offers limited robustness. To enhance resilience, VFLIP~\cite{cho2024vflip} introduces identification and purification strategies. While it achieves partial mitigation, experimental results show that backdoor threats are not fully eliminated.

 \noindent
\textbf{Our work: }%
To strengthen resilience against backdoor attacks in SL, we introduce \alg, a two-step defense strategy tailored to identify and neutralize poisoned embeddings in SL frameworks.
Our \alg leverages an innovative embedding transformation pipeline combined with an adaptive majority-based filtering technique, strategically designed to enhance the separability between benign and poisoned embeddings. By systematically manipulating the embedding space, \alg effectively identifies subtle yet critical discrepancies from adversarial manipulation, thereby significantly improving robustness against stealthy backdoor attacks.

In the first step, we implement a carefully structured dimensionality transformation approach composed of dimensionality reduction followed by dimensional expansion. Specifically, we first apply a dimensionality reduction technique to compress the original high-dimensional embedding space into a lower-dimensional representation. This step removes noisy dimensions while preserving local geometric relationships, ensuring that structural proximity between benign and poisoned embeddings remains intact and discernible. However, as dimensionality reduction might result in a loss of nuanced information crucial for distinguishing benign from poisoned embeddings, we subsequently apply a dimensional expansion method. This transformation reintroduces non-linear feature interactions, amplifying subtle differences and substantially enhancing separability.

Following the embedding transformation, our second step introduces an adaptive majority-based filtering method aimed at robustly identifying poisoned embeddings. Initially, we compute the coordinate-wise median of the transformed embeddings, establishing a robust central reference point indicative of benign embedding behavior. We then employ a majority-based radius determination that retains at least half of the embeddings closest to this median. To address potential misclassification issues and improve precision, we dynamically adjust this radius according to the variance observed within the embeddings: a lower variance leads to a more aggressive enlargement of the radius to capture additional benign embeddings, while higher variance dictates a conservative adjustment to maintain defense robustness. 
This adaptive strategy balances excluding poisoned embeddings and retaining benign ones, improving detection performance.

We rigorously evaluate our proposed \alg across four diverse datasets (i.e., CIFAR-10~\cite{cifar10data}, MNIST~\cite{lecun1998gradient}, CINIC-10~\cite{darlow2018cinic}, and ImageNette~\cite{howard2020fastai}), assessing its resilience against five types of backdoor attacks, including a strong adaptive attack. Additionally, we compare its performance against seven state-of-the-art SL baselines, encompassing extensions of standard HFL approaches such as Trimmed-mean~\cite{yin2018byzantine} and Multi-Krum~\cite{blanchard2017machine}, clustering-based methods like HDBSCAN~\cite{mcinnes2017hdbscan}, and defenses specifically designed for the SL framework. The results demonstrate that our \alg effectively enhances robustness and adaptability across a broad spectrum of real-world backdoor scenarios.
The key contributions of this paper are summarized as follows:
\begin{list}{\labelitemi}{\leftmargin=1.15em \itemindent=-0.08em \itemsep=.2em}

\item
We present \alg, an innovative defense designed to counteract backdoor attacks in split learning.

\item
We conduct a thorough empirical evaluation of \alg in the face of various backdoor attacks, with our results confirming its capability to effectively defend against them.

\item
We craft an adaptive attack against \alg and evaluate its impact, revealing that \alg maintains its resilience even under this targeted threat.

\end{list}

\section{Preliminaries and Related Work}
\label{sec:preliminaries}
\subsection{Background on Split Learning}

Collaborative learning frameworks like federated learning (FL) allow multiple clients to train models without sharing raw data. While preserving privacy, approaches differ in data distribution. In horizontal federated learning (HFL)~\cite{webank,paulik2021federated,chen2024contributions,xie2022federatedscope}, each client holds labeled data with a shared feature space.
In contrast, SL~\cite{fu2022blindfl,romanini2021pyvertical,liu2020asymmetrical} operates under a vertically partitioned data setting, where clients hold different subsets of features for the entire training dataset and do not have access to labels, which are exclusively maintained by a central server.
Formally, consider a SL system with \( n \) clients, each denoted as \( C_i \). Let \( \bm{x}_k \) be the complete feature vector of a training example \( k \), represented as
$
\bm{x}_k = [x_{k,1}, x_{k,2}, ..., x_{k,z}]
$,
where \( z \) is the total number of features. In SL, each client \( C_i \) holds only a subset of the features, denoted as \( \hat{\bm{x}}_k^i \). The full feature vector is distributed across all clients such that
$
\bm{x}_k = \bigcup_{i=1}^{n} \hat{\bm{x}}_k^i
$.
Unlike FL, where clients manage both features and labels, in SL, the label \( y_k \) for each training example is stored solely on the server.
In each training round \( t \), SL follows three fundamental steps. 
For clarity, we omit $t$ in the subsequent explanation.

\begin{list}{\labelitemi}{\leftmargin=1.15em \itemindent=-0.08em \itemsep=.2em}

\item {\bf Step I (Feature embeddings generation):} Each client \( C_i \) processes its local feature subset using a bottom model \( \mathcal{L}_i \), generating an embedding vector:  
\begin{align}
    \bm{E}_k^i = \mathcal{L}_i(\hat{\bm{x}}_k^i),
\end{align}
where \( \bm{E}_k^i \) is the embedding for training example \( k \) from client \( C_i \), $d$ is the dimension of $\bm{E}_k^i$. Clients then send their embeddings to the server for further processing.

\item {\bf Step II (Embeddings aggregation):} Upon receiving embedding vectors from the clients, the server applies an aggregation function \( \mathcal{A}(\cdot) \) to merge them as follows:
\begin{align}
\label{embed_agg}
\bm{E}_k = \mathcal{A}(\bm{E}_k^1, \bm{E}_k^2, ..., \bm{E}_k^n).
\end{align}
This aggregated embedding \( \bm{E}_k  \in \mathbb{R}^d \), along with its corresponding label \( y_k \), is then used to train a top model in a supervised learning framework. Once the model updates are computed, the server calculates the gradient \( \bm{g}_i \) of the loss with respect to each client’s embedding \( \bm{E}_k^i \). Finally, the server transmits \( \bm{g}_i \) back to the respective client \( C_i \) for updating their bottom models.

\item {\bf Step III (Bottom models updating):} Each client \( C_i \) updates the parameters of its bottom model \( \mathcal{L}^i \) using the received gradient \( \bm{g}_i \). 
This process allows each client to refine its bottom model using server feedback, leading to improved feature representations in future rounds.

\end{list}

Note that for simplicity, we assume that each client transmits the embedding of a single training example to the server in the three steps outlined above. However, in practice, clients in SL send embeddings of multiple training examples to the server, which can be seamlessly extended from the single-example transmission process.
We note that SL differs fundamentally from both horizontal federated learning (HFL) and vertical federated learning (VFL), see Appendix~\ref{app_fl_diff} for details.

\subsection{Backdoor Attacks to SL}

SL is a decentralized learning framework where clients share embeddings with the server during each training round. However, this collaborative structure and frequent communication between clients and the server create vulnerabilities, making SL susceptible to backdoor attacks~\cite{he2023backdoor,bai2023villain,naseri2024badvfl}. These attacks enable the attacker to manipulate the learned top model, causing it to produce attacker-specified outputs for selected inputs.
To execute a backdoor attack, a malicious client can introduce hidden triggers into its training data, subtly poisoning the learning process. For example, in BadVFL attack~\cite{naseri2024badvfl}, the attacker embeds triggers directly into the local datasets of malicious clients, influencing the top model’s behavior. Beyond data-level poisoning, the attacker can manipulate embeddings to stealthily implant backdoors. The VILLAIN attack~\cite{bai2023villain} exemplifies this strategy, where the attacker crafts malicious embeddings while fine-tuning the trigger's magnitude to maximize impact while minimizing detectability.

\subsection{Defenses against Backdoor Attacks to SL}
To defend against backdoor attacks in SL, a common strategy is to adapt Byzantine-robust aggregation from FL~\cite{yin2018byzantine,blanchard2017machine,fang2024byzantine,fang2025we,cao2020fltrust,fang2025provably,fang2025byzantine,wang2025poisoning,xie2024fedredefense,dou2025toward,fang2022aflguard,zhang2024securing,mo2025find}. 
For example, Trimmed-mean~\cite{yin2018byzantine} removes extreme values in each embedding dimension before averaging, while clustering-based methods like HDBSCAN~\cite{mcinnes2017hdbscan} group embeddings to filter out potential threats. However, as shown in our experiments, these methods are largely ineffective against sophisticated SL-specific attacks.
More recent defenses include VFLIP~\cite{naseri2024badvfl}, which uses identification and purification techniques but suffers from high false positives, often misclassifying benign embeddings. 
Another recent work~\cite{rieger2025safesplit} proposed a defense specifically designed for U-shaped SL, where clients are trained sequentially. 
Note that here we mean the training order among clients rather than the operation order between the clients and the server.
As a result, the method in~\cite{rieger2025safesplit} is not applicable to our standard SL setting, where clients train in parallel. In Section~\ref{Experimental_results}, we adapt our method to the U-shaped SL setting and compare it with~\cite{rieger2025safesplit}; the results are shown in Table~\ref{U_results} in the Appendix.

\section{Problem Statement}

\begin{figure}[t]
\centering
\includegraphics[scale = 0.3]{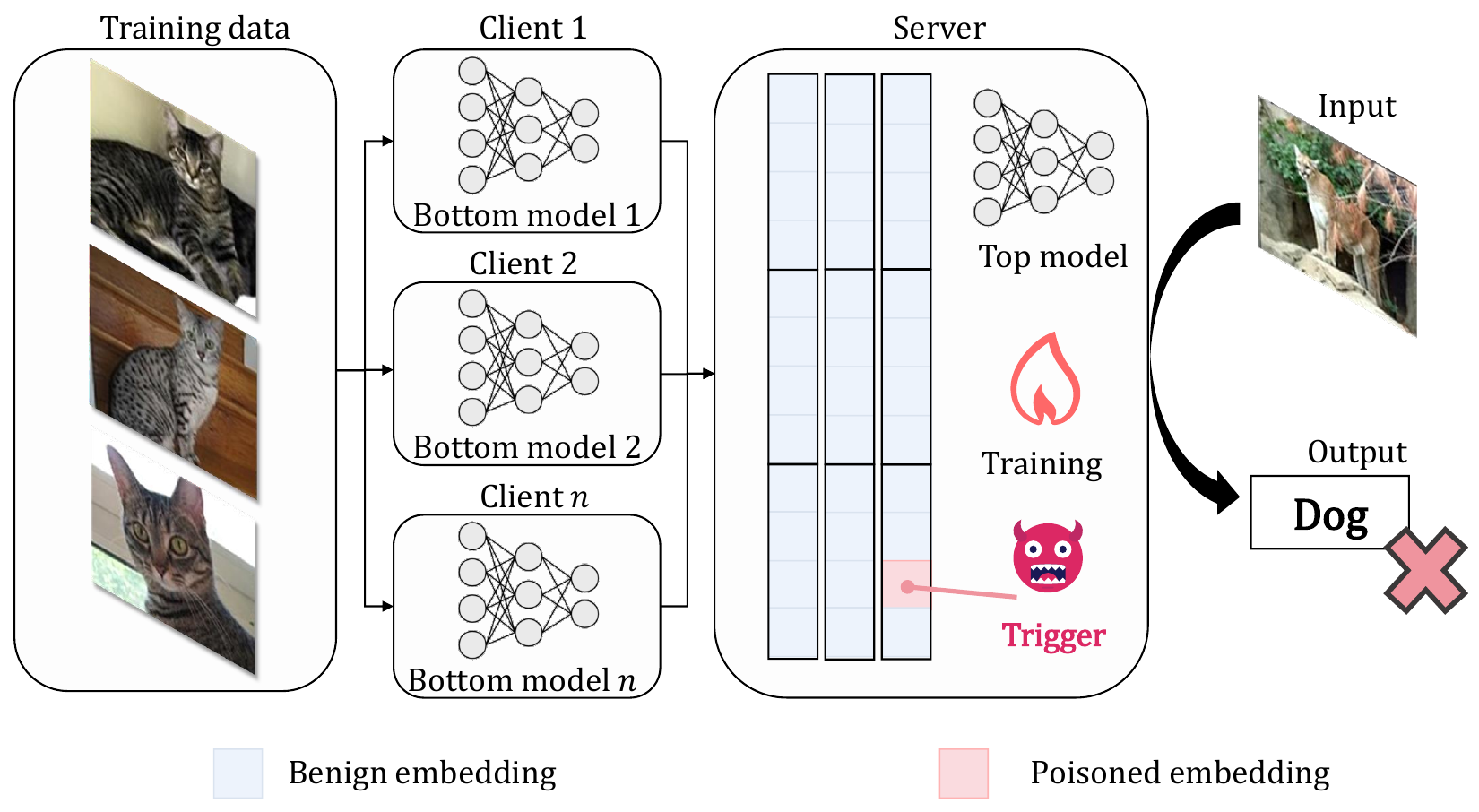}
\caption{Split learning under backdoor attack.}
\label{fig:graph}
 \vspace{-.1in}
\end{figure}

\myparatight{Threat model}%
Building on the threat model from prior work~\cite{he2023backdoor,bai2023villain,fu2022label}, we assume an attacker controlling one or more malicious clients. These clients can introduce backdoor triggers by poisoning local data or embedding manipulated triggers into their transmitted embeddings. As shown in Fig.~\ref{fig:graph}, such embeddings cause the top model to produce attacker-specified outputs.
Consistent with existing SL backdoor attacks~\cite{he2023backdoor,bai2023villain,naseri2024badvfl}, we consider a sophisticated attacker with access to limited labeled auxiliary data, enabling them to infer training labels and craft more precise attacks.
Note that in our threat model, we assume the majority of training data are benign. This assumption is widely adopted in the machine learning and FL communities~\cite{fang2020local}, as the attacker aims to remain stealthy. 
If most training data were malicious, the attacker would be easily detected, violating the covert-adversary assumption.

\myparatight{Defender’s knowledge and goal}%
Our goal is to design a robust defense against backdoor attacks in SL, despite the server receiving only client embeddings without access to their data, models, or attack strategies. Such a defense must minimally affecting model performance when no attacks occur, and effectively identifying and excluding poisoned embeddings to maintain high accuracy and drastically reduce attack success rates.


\section{The \alg Algorithm} 
\label{sec:alg}

\begin{figure}[t]
    \centering
    \includegraphics[scale = 0.3]{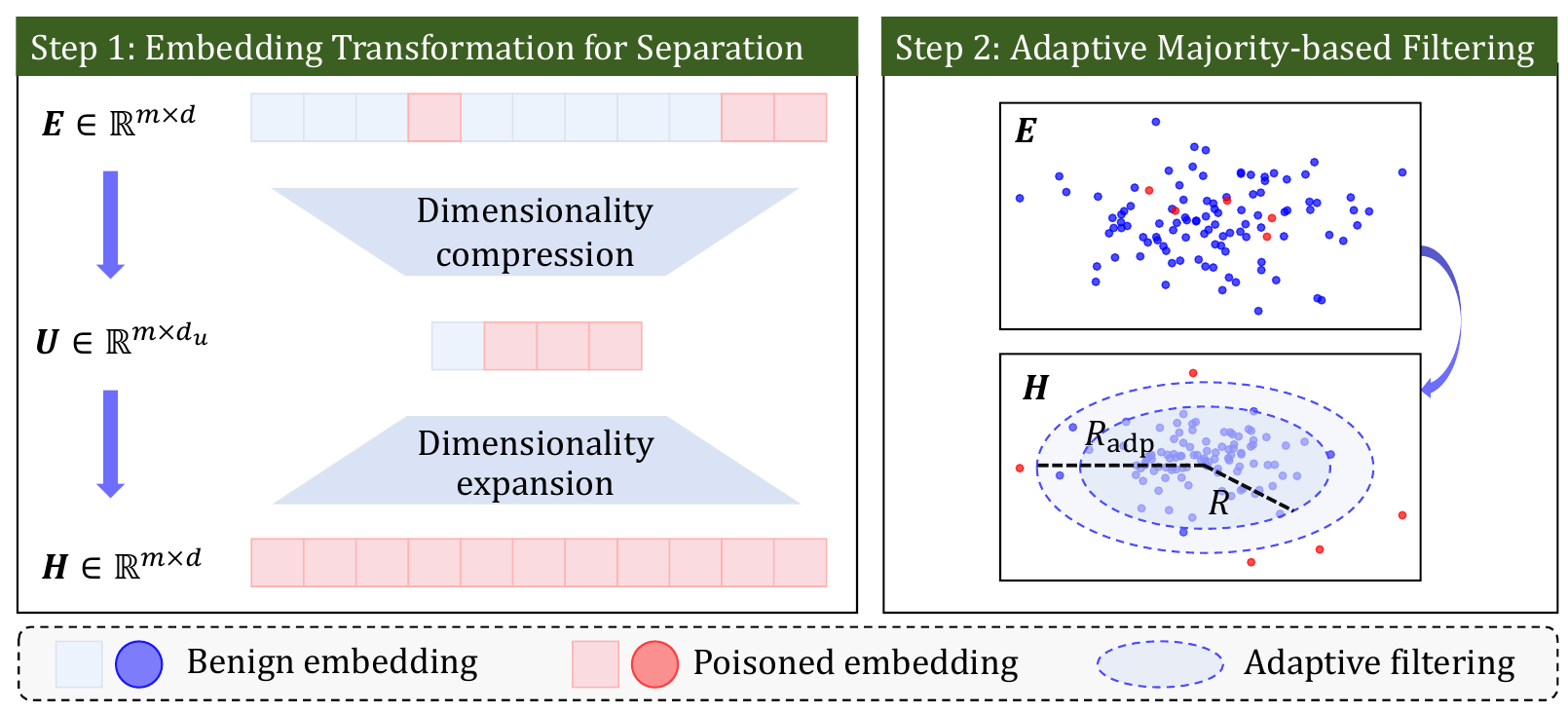}
    \caption{Illustration of our \alg.}
    \label{fig:framework}
 \vspace{-.1in}
\end{figure}

\subsection{Overview}

Our \alg enhances the robustness of SL against backdoor attacks through two main steps: embedding transformation and adaptive filtering. The transformation restructures the embedding space to amplify subtle differences introduced by backdoor triggers. The adaptive filter then removes suspicious embeddings based on majority consensus, dynamically adjusting its criteria to reduce misclassification. Together, these steps effectively mitigate backdoor attacks while preserving model performance.
Fig.~\ref{fig:framework} illustrates the overall framework. We next detail how \alg operates in each training round.

\subsection{Embedding Transformation for Separation}

Consider a dataset consisting of \( m \) training examples, where each example is represented by a feature embedding. We denote the full set of embeddings as \( \bm{E}=\{\bm{E}_1,\bm{E}_2,...,\bm{E}_m\} \), where each individual embedding \( \bm{E}_k \) (for \( k=1,2,...,m \)) is obtained according to Eq.~(\ref{embed_agg}). Since each embedding is a \( d \)-dimensional vector, we have \( \bm{E} \in \mathbb{R}^{m \times d} \). 
In a SL setting, backdoor attacks involve injecting triggers into the embeddings of malicious clients, altering their representation. As a result, \( \bm{E} \) may contain both benign embeddings (\( \bm{E}_{\text{Ben}} \)) and poisoned embeddings (\( \bm{E}_{\text{Poi}} \)). While the modifications introduced by the attacker are often subtle, they can still alter the spatial distribution of embeddings, shifting the positional relationships between poisoned and benign samples. 

To address this challenge, we first apply dimensionality compression using Uniform Manifold Approximation and Projection (UMAP)~\cite{mcinnes2018umap}, transforming the original embedding space into a lower-dimensional representation. The compressed embedding set \( \bm{U} \in \mathbb{R}^{m \times d_u} \) is obtained as:
\begin{align}
    \label{eq:umap}
    \bm{U} = \text{UMAP}(\bm{E}),
\end{align}
where \( \bm{U} = \{\hat{\bm{E}}_1, \hat{\bm{E}}_2, ..., \hat{\bm{E}}_m\} \), each embedding \( \hat{\bm{E}}_k \) (for \( k=1,2,...,m \)) is now in a reduced-dimensional space \( \mathbb{R}^{d_u} \) with \( d_u < d \). 
The motivation for dimensionality reduction is twofold.
First, UMAP is designed to preserve the local geometric structure of data, ensuring that the positional relationships (e.g., relative proximity) between \( \bm{E}_{\text{Ben}} \) and \( \bm{E}_{\text{Poi}} \) in the original space are maintained in \( \bm{U} \). 
Second, by reducing dimensions, UMAP filters out noisy features that obscure key differences, enhancing the separation of poisoned and benign embeddings while preserving important structural patterns.

However, while dimensionality reduction enhances local relationships, it may also cause a loss of fine-grained information that is useful for separating benign and poisoned embeddings. To compensate for this, we introduce a dimensionality expansion step using Polynomial Kernel Transformation (PKT)~\cite{weisse2006kernel}, which maps the compressed embeddings back to a higher-dimensional space, restoring and amplifying crucial variations. The expanded embedding set \( \bm{H} \in \mathbb{R}^{m \times d} \) is obtained as follows:
\begin{align}
    \label{eq:pk}
    \bm{H} = \text{PKT}(\bm{U}),
\end{align}
where \( \bm{H} = \{\bar{\bm{E}}_1, \bar{\bm{E}}_2, ..., \bar{\bm{E}}_m \} \), with each \( \bar{\bm{E}}_k \) is in the restored high-dimensional space \( \mathbb{R}^{d} \). The PKT expansion introduces nonlinear transformations that enhance the separation between benign and poisoned embeddings by leveraging higher-order feature interactions. This step is crucial as it amplifies subtle distance variations, making anomalies more distinguishable and easier to detect.

In summary, our \alg strategically combines dimensionality reduction and expansion to maximize the separability of poisoned and benign embeddings. First, UMAP removes noise dimensions while retaining the core structural relationships between embeddings. Then, PKT reintroduces higher-dimensional representations that emphasize and enhance these relationships, allowing for a more distinct separation between poisoned and benign embeddings in the final space.
This two-step approach provides a robust mechanism for detecting backdoor attacks by making poisoned embeddings more distinguishable from their benign counterparts.

\begin{figure}[t]
    \centering
    \includegraphics[width=0.75\linewidth]{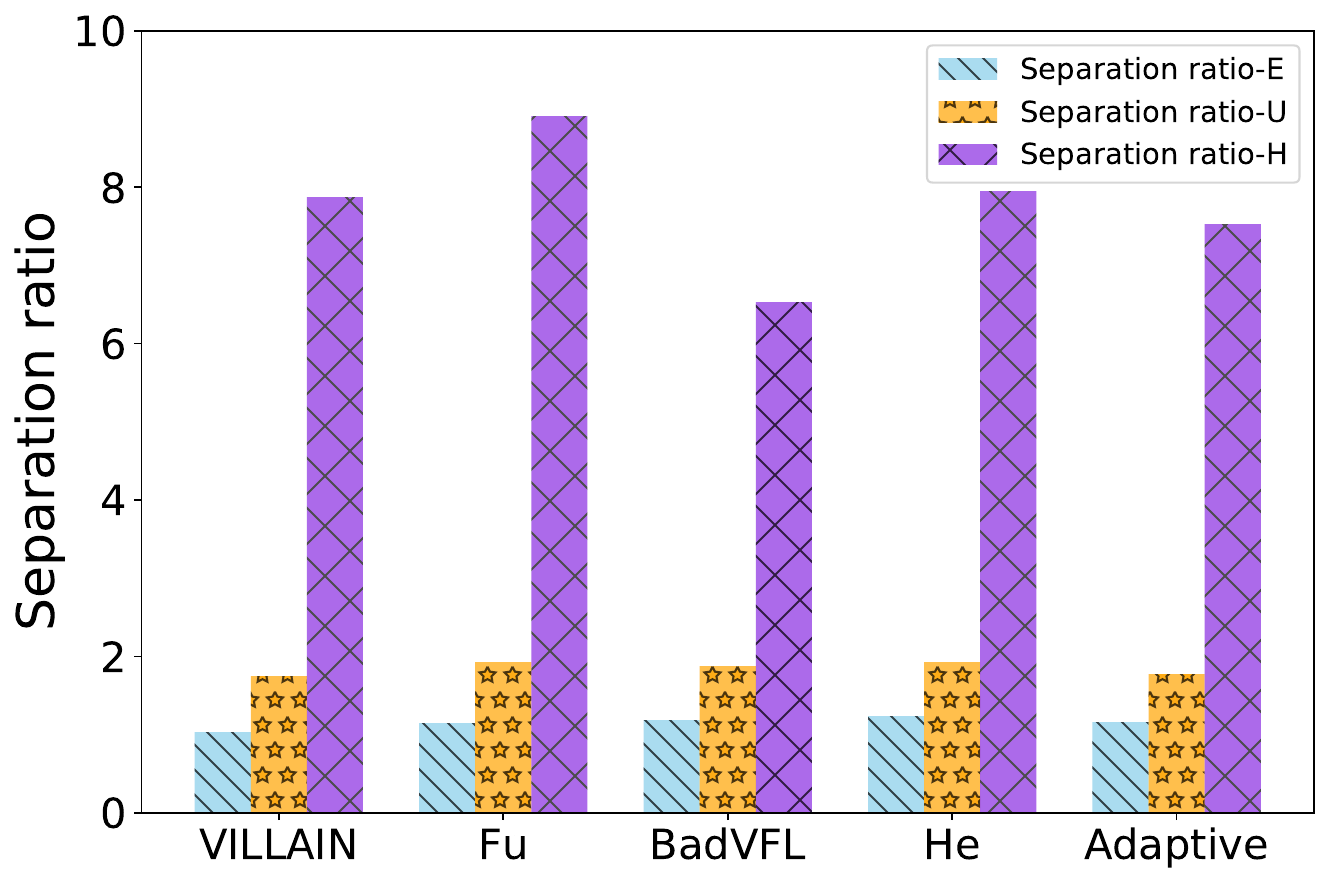}
    \caption{Separation ratio on different embedding sets.}
    \label{DR_ER_results}
 \vspace{-.1in}
\end{figure}

\myparatight{Empirical analysis}%
To validate \alg, we conduct experiments on CIFAR-10, showing that dimensionality reduction and expansion improve the separation between benign and poisoned embeddings. We generate 100 benign-benign and 100 benign-poisoned embedding pairs, computing their average distances as $D_{\text{Ben}}$ and $D_{\text{Poi}}$, respectively. The separation ratio is defined as
$
\text{Separation ratio} = \frac{D_{\text{Poi}}}{D_{\text{Ben}}}.
$
Fig.~\ref{DR_ER_results} reports the separation ratios under five backdoor attacks: VILLAIN~\cite{bai2023villain}, Fu~\cite{fu2022label}, BadVFL~\cite{naseri2024badvfl}, He~\cite{he2023backdoor}, and Adaptive attacks. ``Separation ratio-E'', ``-U'', and ``-H'' refer to calculations on embedding sets $\bm{E}$, $\bm{U}$, and $\bm{H}$. Results show that while reduction yields moderate separation, expansion significantly boosts it, making poisoned embeddings more distinguishable. This highlights the effectiveness of \alg in enhancing embedding discrimination, even under stealthy attacks.
However, as later shown in Table~\ref{various_components}, directly applying clustering to $\bm{H}$ remains ineffective for defense.

\subsection{Adaptive Majority-based Filtering}

Once $\bm{H}$ is obtained, the separation between poisoned and benign embeddings becomes more pronounced. Building on this, we propose an adaptive detection method to identify poisoned embeddings. 
This design is motivated by the observation that backdoor attacks in SL typically poison only a small fraction of embeddings to remain stealthy. Poisoning too many embeddings would not only increase the risk of detection but also introduce excessive manipulation costs, making the attack less feasible. 
Thus, most embeddings remain benign, and poisoning a majority is impractical without distorting the feature space or degrading performance.
Building on this intuition, we design a majority-based filtering mechanism that retains embeddings consistent with the majority consensus. Specifically, the server first computes the coordinate-wise median~\cite{yin2018byzantine} of all embeddings in \( \bm{H} \) as:  
\begin{align}
\bm{\Lambda}=\text{Median}(\bar{\bm{E}}_1, \bar{\bm{E}}_2, ..., \bar{\bm{E}}_m),
\end{align}
where \(\text{Median}(\cdot)\) represents the coordinate-wise median function, producing \(\bm{\Lambda} \in \mathbb{R}^d\).

\begin{algorithm}[t]
    \caption{\alg.}
    \label{algo:securesplit}
    \begin{algorithmic}[1]
        \renewcommand{\algorithmicrequire}{\textbf{Input:}}
        \renewcommand{\algorithmicensure}{\textbf{Output:}}
        \Require Full set of embeddings $\bm{E}$, parameter \(\alpha\).
        \Ensure Benign embedding set $\mathcal{S}$.
        \State $\mathcal{S} \leftarrow \emptyset$.
        \State \textcolor{blue}{// Step I: Embedding Transformation for Separation.}
        \State Compute compressed embeddings $\bm{U}$ using Eq.~(\ref{eq:umap}).
        \label{step_one_one}
        \State Compute expanded embeddings $\bm{H}$ per Eq.~(\ref{eq:pk}).
        \label{step_one_two}
        \State \textcolor{blue}{// Step II: Adaptive Majority-based Filtering.}
        \State Calculate coordinate-wise median: $\bm{\Lambda} \gets \text{Median}(\bm{H})$.
         \label{step_two_one}
        \State Determine radius $R$ using Eq.~(\ref{compute_R}).
        \State Compute variance $\sigma$ of embeddings in $\bm{H}$.
        \State Calculate adaptive radius $R_{\text{adp}}$ according to Eq.~(\ref{r_opt}).
        \For{each $\bar{\bm{E}}_k$ in $\bm{H}$}
        \If {Eq.~(\ref{our_filter}) is satisfied}
         \State $\mathcal{S} \leftarrow \mathcal{S} \bigcup \{\bar{\bm{E}}_k \} $.
        \EndIf
        \EndFor
        \label{step_two_two}
    \end{algorithmic}
\end{algorithm}

Next, we determine whether an embedding \( \bar{\bm{E}}_k \) is likely benign by evaluating its closeness to \(\bm{\Lambda}\). An embedding is considered potentially benign if it satisfies the condition  
$
\left\| \bar{\bm{E}}_k -  \bm{\Lambda} \right\|_2 \le R,
$
where \( R \) is a radius parameter, $\left\| \cdot \right\|_2 $ is the $\ell_2$ norm. Conceptually, \(\bm{\Lambda} \) serves as the centroid of a hypersphere, and embeddings within a distance \( R \) from it are accepted as benign.  
To determine \( R \), we ensure that at least half of the embeddings lie within this ball by computing:  
\begin{align}
\label{compute_R}
R = \min \left\{ R' \mid \left| \left\{ k \in [m]: \left\| \bar{\bm{E}}_k - \bm{\Lambda} \right\|_2 \le R' \right\} \right| \ge \frac{m}{2} \right\}, 
\end{align}
where \( [m] \) denotes the set \(\{1,2,\dots,m\}\), and \( \left| \cdot \right| \) represents the cardinality of a set. Eq.~(\ref{compute_R}) ensures that \( R \) is the smallest possible value that includes at least half of the embeddings.

However, a key limitation of this filtering mechanism is its tendency to misclassify a significant number of benign embeddings as poisoned. Since the selection criterion in Eq.~(\ref{compute_R}) retains only half of the embeddings, a large fraction of benign embeddings may still be excluded. To mitigate this issue, we refine our approach by slightly enlarging \( R \) to incorporate more benign embeddings while still excluding poisoned ones.  
Our key insight is that poisoned embeddings tend to deviate significantly from benign ones, leading to increased variance in \( \bm{H} \) after an attack. Therefore, we introduce an adaptive adjustment to \( R \) based on the variance \( \sigma \) of embeddings in \( \bm{H} \). If \( \sigma \) is small, we increase \( R \) more aggressively to capture additional benign embeddings. Conversely, if \( \sigma \) is large, we apply a more conservative enlargement to maintain robustness against poisoning. The adaptive radius, \( R_{\text{adp}} \), is computed as follows:  
\begin{align}
\label{r_opt}
R_{\text{adp}} =  (1+\frac{1}{\alpha+\sigma}) R,
\end{align}
where \( \alpha \) is a parameter that controls the sensitivity of the adjustment.  
To determine whether an embedding $\bar{\bm{E}}_k$ is benign, we assess its proximity to $\bm{\Lambda}$. If it meets the following condition, it is classified as benign:
\begin{align}
\label{our_filter}
\left\| \bar{\bm{E}}_k -  \bm{\Lambda} \right\|_2 \le R_{\text{adp}},
\end{align}  
where \( R_{\text{adp}} \) is adaptively computed using Eq.~(\ref{r_opt}), with \( R \) in Eq.~(\ref{r_opt}) determined by Eq.~(\ref{compute_R}).

By dynamically adapting the radius parameter based on the distribution of embeddings, our method effectively balances the trade-off between excluding poisoned embeddings and preserving benign ones, enhancing robustness against backdoor attacks in SL.
Algorithm~\ref{algo:securesplit} outlines the pseudocode for our \alg. In each training round, once the server receives the complete set of embeddings \( \bm{E} \) from the clients, it employs a two-step approach to mitigate the effects of backdoor attacks. In the first step (Lines~\ref{step_one_one}-\ref{step_one_two} of Algorithm~\ref{algo:securesplit}), the server applies a dimensionality compression and expansion technique to improve the distinguishability of poisoned embeddings. Subsequently, the adaptive majority-based filtering mechanism is used to identify and remove potential malicious embeddings, ultimately producing the benign embedding set \( \mathcal{S} \) for that round (Lines~\ref{step_two_one}-\ref{step_two_two}).


\section{Experimental Evaluation}  
\label{sec:exp}

\begin{table}[]
\centering
\footnotesize
\addtolength{\tabcolsep}{-1.755pt}
\caption{Performance of defense methods is evaluated using ACC (\(\uparrow\)) and ASR (\(\downarrow\)) metrics, where higher ACC and lower ASR indicate better performance.}

\begin{tabular}{|c|c|cc|cc|cc|cc|}
\hline
\multirow{2}{*}{Attack}    & \multirow{2}{*}{Defense} & \multicolumn{2}{c|}{CIFAR-10} & \multicolumn{2}{c|}{MNIST} & \multicolumn{2}{c|}{CINIC-10} & \multicolumn{2}{c|}{ImageNette} \\ \cline{3-10} 
                           &                          & ACC            & ASR          & ACC           & ASR         & ACC            & ASR           & ACC             & ASR            \\ \hline
\multirow{3}{*}{}          & No defense               & 0.84          & 0.76         & 0.97         & 0.92        & 0.65          & 0.73          & 0.71           & 0.77           \\
                           & TrMean                   & 0.83          & 0.57         & 0.91         & 0.72        & 0.55          & 0.49          & 0.54           & 0.57           \\
                           & Multi-Krum               & 0.77          & 0.48         & 0.87         & 0.59        & 0.49          & 0.45          & 0.47           & 0.49           \\
VILLAIN                    & HDBSCAN                  & 0.82          & 0.55         & 0.92         & 0.74        & 0.65          & 0.63          & 0.71           & 0.70           \\
attack                     & DP                       & 0.83          & 0.57         & 0.96         & 0.72        & 0.63          & 0.61          & 0.68           & 0.57           \\
\multirow{4}{*}{}          & MP                       & 0.80          & 0.67         & 0.94         & 0.75        & 0.60          & 0.62          & 0.70           & 0.64           \\
                           & ANP                      & 0.82          & 0.48         & 0.95         & 0.62        & 0.64          & 0.58          & 0.65           & 0.52           \\
                           & VFLIP                    & 0.76          & 0.17         & 0.94         & 0.08        & 0.58          & 0.16          & 0.64           & 0.15           \\
                           & \cellcolor{greyL} \alg                     & \cellcolor{greyL} 0.85          & \cellcolor{greyL} 0.06         & \cellcolor{greyL}0.96         & \cellcolor{greyL}0.02        & \cellcolor{greyL}0.67          & \cellcolor{greyL}0.04          & \cellcolor{greyL}0.74           & \cellcolor{greyL}0.06           \\ \hline
\multirow{9}{*}{Fu attack} & No defense               & 0.82          & 0.72         & 0.97         & 0.94        & 0.62          & 0.67          & 0.68           & 0.72           \\
                           & TrMean                   & 0.74          & 0.62         & 0.88         & 0.63        & 0.57          & 0.42          & 0.64           & 0.47           \\
                           & Multi-Krum               & 0.67          & 0.65         & 0.82         & 0.55        & 0.55          & 0.40          & 0.61           & 0.44           \\
                           & HDBSCAN                  & 0.75          & 0.66         & 0.89         & 0.80        & 0.64          & 0.55          & 0.69           & 0.64           \\
                           & DP                       & 0.82          & 0.48         & 0.95         & 0.92        & 0.66          & 0.38          & 0.66           & 0.50           \\
                           & MP                       & 0.82          & 0.48         & 0.97         & 0.62        & 0.63          & 0.39          & 0.70           & 0.49           \\
                           & ANP                      & 0.80          & 0.33         & 0.94         & 0.48        & 0.62          & 0.31          & 0.67           & 0.39           \\
                           & VFLIP                    & 0.74          & 0.09         & 0.92         & 0.07        & 0.58          & 0.13          & 0.62           & 0.11           \\
                           & \cellcolor{greyL}\alg                     & \cellcolor{greyL}0.85          & \cellcolor{greyL}0.05         & \cellcolor{greyL}0.98         & \cellcolor{greyL}0.03        & \cellcolor{greyL}0.66          & \cellcolor{greyL}0.07          & \cellcolor{greyL}0.73           & \cellcolor{greyL}0.07           \\ \hline
\multirow{3}{*}{}          & No defense               & 0.80          & 0.47         & 0.94         & 0.70        & 0.67          & 0.46          & 0.64           & 0.43           \\
                           & TrMean                   & 0.77          & 0.40         & 0.88         & 0.42        & 0.60          & 0.42          & 0.66           & 0.45           \\
                           & Multi-Krum               & 0.64          & 0.34         & 0.91         & 0.33        & 0.52          & 0.44          & 0.52           & 0.41           \\
BadVFL                     & HDBSCAN                  & 0.80          & 0.44         & 0.93         & 0.59        & 0.64          & 0.42          & 0.71           & 0.65           \\
attack                     & DP                       & 0.82          & 0.35         & 0.96         & 0.35        & 0.66          & 0.43          & 0.71           & 0.57           \\
\multirow{4}{*}{}          & MP                       & 0.80          & 0.42         & 0.94         & 0.49        & 0.67          & 0.39          & 0.69           & 0.62           \\
                           & ANP                      & 0.78          & 0.30         & 0.94         & 0.17        & 0.64          & 0.37          & 0.65           & 0.48           \\
                           & VFLIP                    & 0.76          & 0.13         & 0.92         & 0.11        & 0.63          & 0.27          & 0.69           & 0.08           \\
                           & \cellcolor{greyL}\alg                     & \cellcolor{greyL}0.85          & \cellcolor{greyL}0.04         & \cellcolor{greyL}0.97         & \cellcolor{greyL}0.03        & \cellcolor{greyL}0.70          & \cellcolor{greyL}0.09          & \cellcolor{greyL}0.75           & \cellcolor{greyL}0.03           \\ \hline
\multirow{9}{*}{He attack} & No defense               & 0.83          & 0.72         & 0.93         & 0.89        & 0.63          & 0.70          & 0.70           & 0.65           \\
                           & TrMean                   & 0.78          & 0.60         & 0.89         & 0.64        & 0.61          & 0.45          & 0.62           & 0.53           \\
                           & Multi-Krum               & 0.70          & 0.58         & 0.82         & 0.52        & 0.57          & 0.38          & 0.58           & 0.44           \\
                           & HDBSCAN                  & 0.80          & 0.69         & 0.94         & 0.75        & 0.63          & 0.52          & 0.70           & 0.62           \\
                           & DP                       & 0.82          & 0.62         & 0.94         & 0.51        & 0.64          & 0.49          & 0.71           & 0.48           \\
                           & MP                       & 0.74          & 0.66         & 0.92         & 0.57        & 0.60          & 0.44          & 0.70           & 0.54           \\
                           & ANP                      & 0.77          & 0.57         & 0.87         & 0.39        & 0.62          & 0.42          & 0.64           & 0.43           \\
                           & VFLIP                    & 0.73          & 0.09         & 0.85         & 0.12        & 0.60          & 0.11          & 0.62           & 0.15           \\
                           & \cellcolor{greyL}\alg                     & \cellcolor{greyL}0.83          & \cellcolor{greyL}0.07         & \cellcolor{greyL}0.97         & \cellcolor{greyL}0.04        & \cellcolor{greyL}0.66          & \cellcolor{greyL}0.03          & \cellcolor{greyL}0.72           & \cellcolor{greyL}0.07           \\ \hline
\multirow{3}{*}{}          & No defense               & 0.83          & 0.79         & 0.94         & 0.94        & 0.65          & 0.78          & 0.73           & 0.78           \\
                           & TrMean                   & 0.74          & 0.72         & 0.87         & 0.90        & 0.63          & 0.72          & 0.64           & 0.59           \\
                           & Multi-Krum               & 0.71          & 0.55         & 0.84         & 0.65        & 0.62          & 0.51          & 0.62           & 0.53           \\
Adaptive                   & HDBSCAN                  & 0.77          & 0.72         & 0.92         & 0.94        & 0.64          & 0.67          & 0.67           & 0.74           \\
attack                     & DP                       & 0.80          & 0.66         & 0.90         & 0.94        & 0.65          & 0.70          & 0.69           & 0.59           \\
\multirow{4}{*}{}          & MP                       & 0.75          & 0.69         & 0.89         & 0.81        & 0.64          & 0.68          & 0.67           & 0.74           \\
                           & ANP                      & 0.79          & 0.58         & 0.92         & 0.68        & 0.65          & 0.66          & 0.67           & 0.59           \\
                           & VFLIP                    & 0.76          & 0.15         & 0.90         & 0.19        & 0.69          & 0.32          & 0.67           & 0.20           \\
                           & \cellcolor{greyL}\alg                     & \cellcolor{greyL}0.87          & \cellcolor{greyL}0.08         & \cellcolor{greyL}0.97         & \cellcolor{greyL}0.05        & \cellcolor{greyL}0.67          & \cellcolor{greyL}0.07          & \cellcolor{greyL}0.73           & \cellcolor{greyL}0.07           \\ \hline
\end{tabular}
\label{main_reuslts}
 \vspace{-0.2cm}
\end{table}

\begin{figure*}
    \centering
    \includegraphics[width=1.0\linewidth]{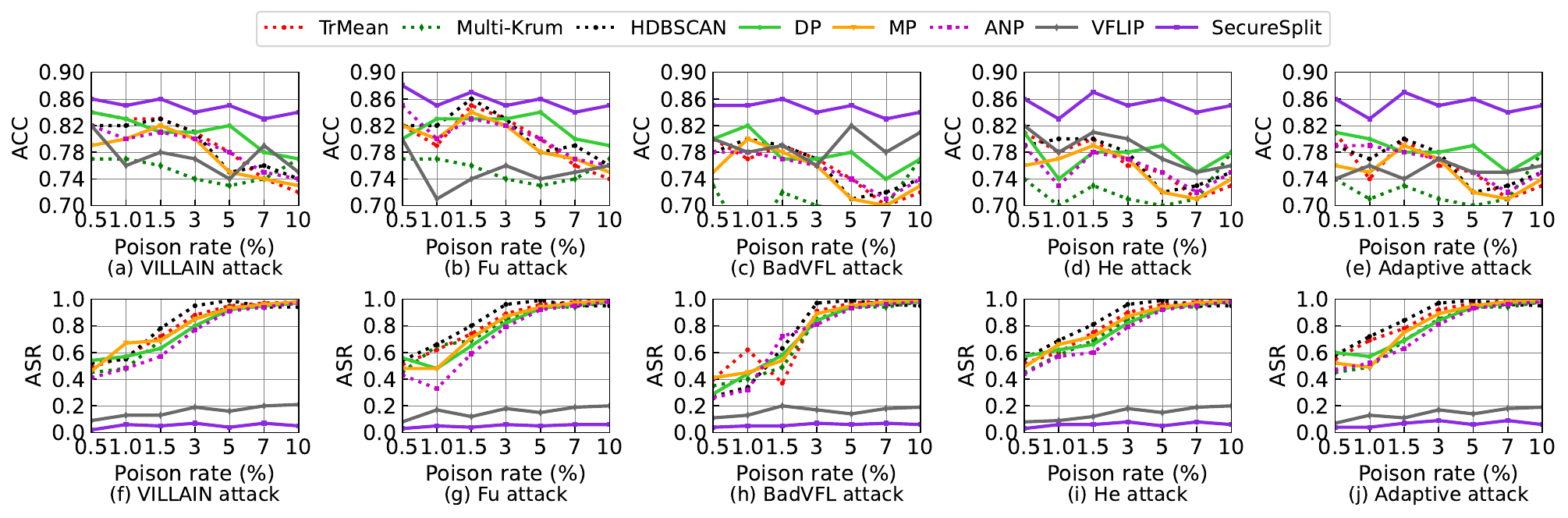} 
    \caption{Impact of the poison rate, where CIFAR-10 dataset is considered.}
    \label{poison_rate}
           \vspace{-0.25cm}
\end{figure*}

\begin{figure*}
    \centering
    \includegraphics[width=1.0\linewidth]{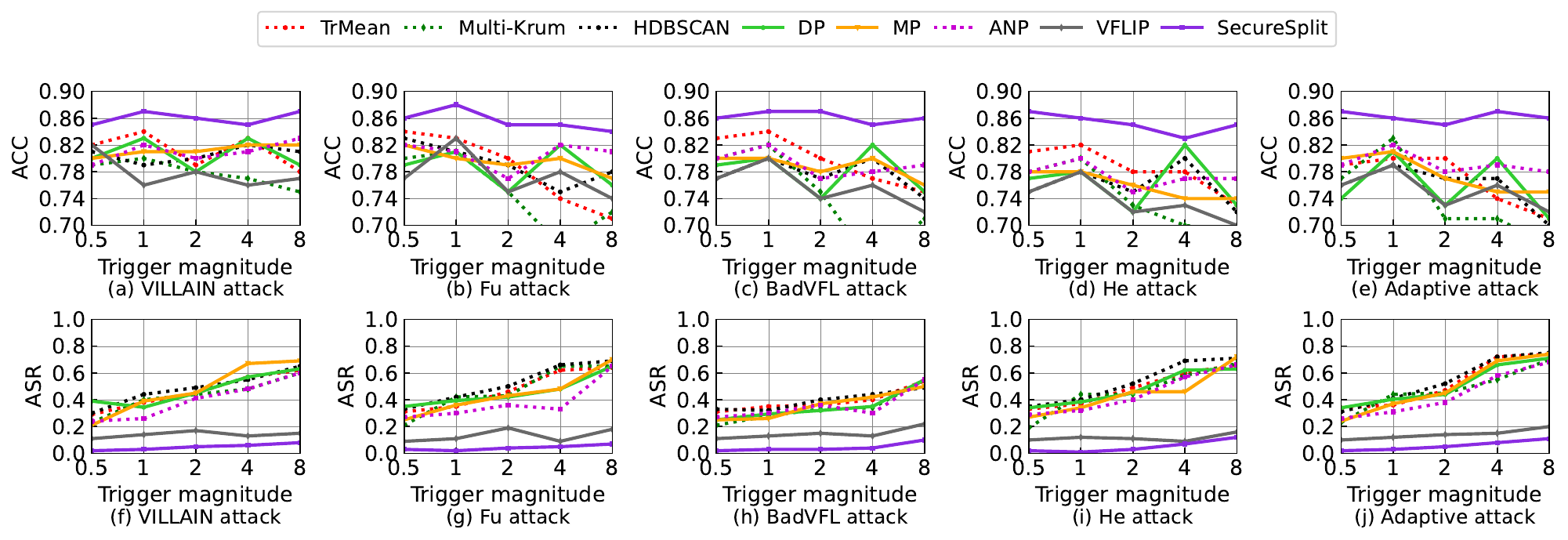} 
    \caption{Impact of the trigger magnitude, where CIFAR-10 dataset is considered.}
    \label{trigge_magnitude}
      \vspace{-0.25cm}
\end{figure*}

\subsection{Experimental Settings}

\myparatight{Datasets}%
We conducted extensive evaluations of \alg across four datasets: CIFAR-10~\cite{cifar10data}, MNIST~\cite{mnist}, CINIC-10~\cite{darlow2018cinic}, and ImageNette~\cite{howard2020fastai} (Appendix~\ref{app_dataset} for more details).

\myparatight{Backdoor attacks}%
We examine five distinct types of backdoor poisoning attacks, comprising four previously established attacks (VILLAIN attack~\cite{bai2023villain}, Fu attack~~\cite{fu2022label}, BadVFL attack~\cite{naseri2024badvfl}, and He attack~\cite{he2023backdoor}), along with a newly designed Adaptive attack.
Appendix~\ref{app_attack} provides an in-depth overview of these backdoor attacks.

\myparatight{Compared methods}%
We benchmark \alg against seven  defenses: Trimmed-mean (TrMean)~\cite{yin2018byzantine}, Multi-Krum~\cite{blanchard2017machine}, HDBSCAN~\cite{malzer2020hybrid}, differential privacy (DP)~\cite{abadi2016deep}, model pruning (MP)~\cite{liu2018fine}, adversarial neuron pruning (ANP)~\cite{wu2021adversarial}, and VFLIP~\cite{cho2024vflip}. Appendix~\ref{app_method} contains further details.

\myparatight{Evaluation metrics}%
Two metrics are used: testing accuracy (ACC) and attack success rate (ASR). ACC represents the proportion of clean testing examples that are correctly classified, while ASR indicates the fraction of trigger-infected testing inputs predicted as the attacker's chosen target label. A higher ACC and a lower ASR indicate better defense performance.

\myparatight{Non-IID setting}%
In SL, each client holds only a portion of the features for the entire training dataset. To simulate feature-level Non-IID heterogeneity, we allocate to each client a subset of features consisting of both shared and unique parts, regulated by an overlapping degree $\rho$. 
Adjusting $\rho$ from 1 to 0 allows us to control the degree of heterogeneity, ranging from fully IID ($\rho = 1$) to the most extreme Non-IID ($\rho = 0$) setting. In our experiments, we adopt the most heterogeneous case by setting $\rho = 0$, meaning clients have entirely disjoint feature subsets. 
See Appendix~\ref{app_noniid} for details.

\myparatight{Parameter setting}%
We assess \alg in a setting with four participating clients, including one malicious client as in~\cite{cho2024vflip}.
Following existing studies~\cite{he2023backdoor,bai2023villain,fu2022label,cho2024vflip}, we randomly partition the feature space vertically across clients, ensuring no overlap among client features.
By default, all clients participate in each training round.
Details on the neural network architectures, learning rates, batch sizes, and total training rounds for each dataset are provided in Appendix~\ref{app_more_settings}.
For embedding aggregation, the concatenation method is used by default, and the reduced dimension $d_u$ is set to 2. To ensure a fair comparison, the trigger size for the VILLAIN and BadVFL attacks aligns with~\cite{naseri2024badvfl}, while the trigger size for the He attack follows the parameters in~\cite{he2023backdoor}. In the Fu attack~\cite{fu2022label}, all poisoned local client embeddings are replaced with the trigger. Across all datasets, the trigger magnitude \(\lambda\) (the coefficient of the backdoor trigger injected into the embedding layer) is set to 4, and the poison rate (the proportion of poisoned embeddings relative to the total training embeddings) is maintained at 1\% following~\cite{bai2023villain}.

\subsection{Experimental results}
\label{Experimental_results}

\myparatight{\alg is effective}%
In Table \ref{main_reuslts}, we evaluate the performance of our proposed \alg method alongside various defense strategies under different attack scenarios.
Our results demonstrate that \alg consistently outperforms other defense methods across multiple datasets and attack types, achieving the highest ACC and the lowest ASR in nearly all cases. For instance, under the VILLAIN attack on CIFAR-10, \alg achieves an ACC of 0.85 and an ASR of 0.06, significantly surpassing methods like TrMean (ACC: 0.83, ASR: 0.57) and Multi-Krum (ACC: 0.77, ASR: 0.48). 
These results highlight the robustness of \alg in mitigating adversarial attacks while preserving model accuracy, making it a superior defense mechanism for secure SL.

\myparatight{Impact of poison rate}%
Fig.~\ref{poison_rate} displays the ACC and ASR results for various defense baselines under five different attacks on the CIFAR-10 dataset, as the poisoned embedding rate increases from 0.5\% to 10\%. As the poison rate rises, the ASR of other defense methods steadily increases, with MP surpassing 95\% ASR when the poison rate reaches 3\%. In contrast, \alg demonstrates exceptional robustness, keeping the ASR below 10\% across all poison rates.

\myparatight{Impact of trigger magnitude}%
Fig.~\ref{trigge_magnitude} illustrates the impact of trigger magnitude on the defense model, with values tested in the range of [0.5, 8]. As shown, the ASR of other methods rises with an increase in trigger magnitude. Under the He attack, when the trigger magnitude reaches 8.0, the ASR of DP approaches 85\%. In contrast, our method consistently keeps the ASR below 0.10 for all tested magnitudes, highlighting its robustness.

\begin{figure*}
    \centering
    \includegraphics[width=1.0\linewidth]{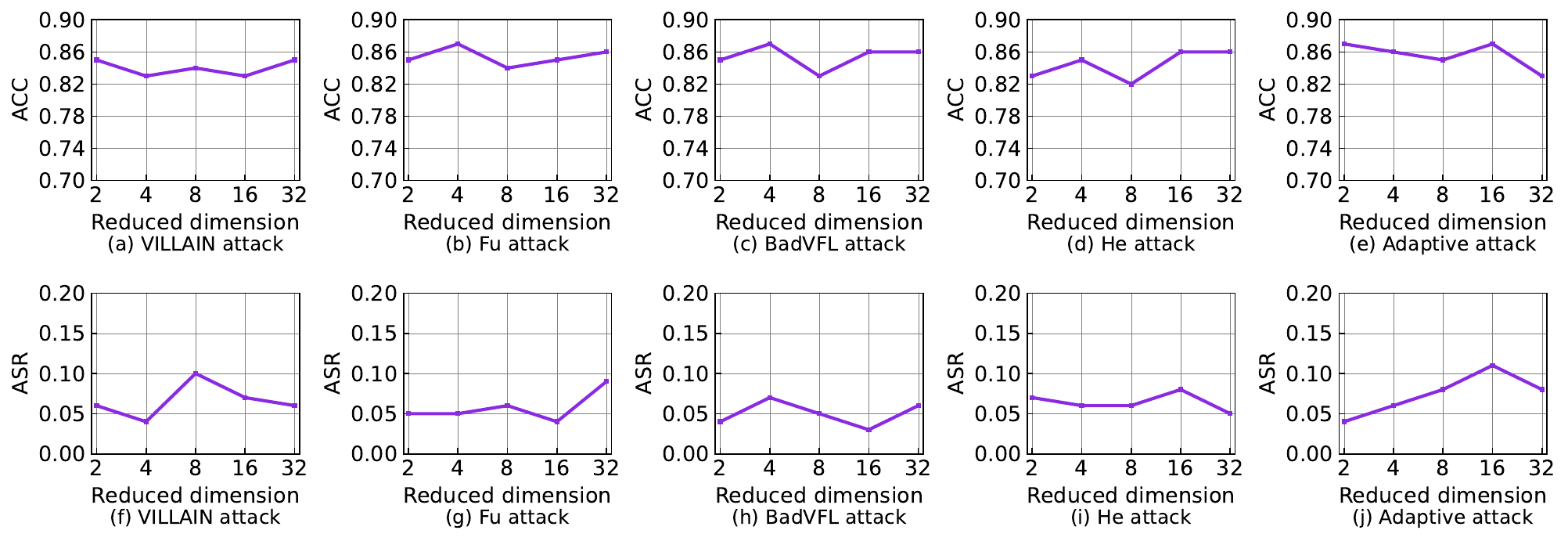} %
    \caption{Impact of the reduced dimension, where the CIFAR-10 dataset is considered.}
    \label{rd}
      \vspace{-0.2cm}
\end{figure*}

\myparatight{Impact of reduced dimension $d_u$}%
Fig.~\ref{rd} shows the effect of different levels of dimensionality reduction on ACC and ASR. Increasing the reduction dimension has minimal impact on the VILLAIN, Fu, and BadVFL attacks, but it negatively influences performance under He and Adaptive attacks.

\begin{table}[t]
\centering
\footnotesize
 \addtolength{\tabcolsep}{-3.75pt} 
\caption{Impact of different embedding aggregation approaches. The CIFAR-10 dataset is considered.}
\begin{tabular}{|c|cc|cc|cc|cc|cc|}
\hline
\multirow{2}{*}{Aggregation} & \multicolumn{2}{c|}{VILLAIN attack} & \multicolumn{2}{c|}{Fu attack} & \multicolumn{2}{c|}{BadVFL attack} & \multicolumn{2}{c|}{He attack} & \multicolumn{2}{c|}{Adaptive attack} \\ \cline{2-11} 
                             & ACC               & ASR              & ACC             & ASR           & ACC               & ASR             & ACC             & ASR           & ACC                & ASR              \\ \hline
Average                      & 0.82             & 0.07             & 0.83           & 0.05          & 0.82             & 0.07            & 0.83           & 0.03          & 0.83              & 0.04             \\
Max                          & 0.87             & 0.04             & 0.82           & 0.04          & 0.83             & 0.08            & 0.84           & 0.06          & 0.86              & 0.06             \\
Min                          & 0.83             & 0.03             & 0.84           & 0.06          & 0.80             & 0.08            & 0.83           & 0.07          & 0.83              & 0.05             \\
Median                       & 0.85             & 0.09             & 0.81           & 0.03          & 0.84             & 0.09            & 0.84           & 0.03          & 0.84              & 0.06             \\
Concatenate                  & 0.85             & 0.06             & 0.85           & 0.05          & 0.85             & 0.04            & 0.83           & 0.07          & 0.87              & 0.08             \\ \hline
\end{tabular}
\label{aggre}
 \vspace{-.05in}
\end{table}

\begin{table}[t]
\footnotesize
 \addtolength{\tabcolsep}{-3.15pt} 
\caption{Performance of various defenses under multiple malicious clients. The CIFAR-10 dataset is considered.}
\centering
\begin{tabular}{|c|cc|cc|cc|cc|cc|}
\hline
\multirow{2}{*}{Defense} & \multicolumn{2}{c|}{VILLAIN attack} & \multicolumn{2}{c|}{Fu attack} & \multicolumn{2}{c|}{BadVFL} & \multicolumn{2}{c|}{He attack} & \multicolumn{2}{c|}{Adaptive attack} \\ \cline{2-11} 
                         & ACC            & ASR          & ACC             & ASR           & ACC           & ASR          & ACC             & ASR           & ACC                & ASR              \\ \hline
No defense               & 0.78          & 0.97         & 0.75           & 0.92          & 0.80         & 0.87         & 0.74           & 0.99          & 0.78              & 0.97             \\
TrMean                   & 0.82          & 0.86         & 0.82           & 0.95          & 0.83         & 0.82         & 0.76           & 0.93          & 0.82              & 0.90             \\
Multi-Krum               & 0.72          & 0.82         & 0.77           & 0.88          & 0.70         & 0.79         & 0.68           & 0.88          & 0.84              & 0.89             \\
HDBSCAN                  & 0.80          & 0.94         & 0.80           & 0.83          & 0.76         & 0.88         & 0.78           & 0.98          & 0.80              & 0.96             \\
DP                       & 0.78          & 0.89         & 0.79           & 0.82          & 0.83         & 0.85         & 0.77           & 0.94          & 0.77              & 0.94             \\
MP                       & 0.79          & 0.85         & 0.82           & 0.84          & 0.79         & 0.78         & 0.80           & 0.94          & 0.81              & 0.96             \\
ANP                      & 0.72          & 0.88         & 0.78           & 0.87          & 0.81         & 0.73         & 0.78           & 0.92          & 0.82              & 0.92             \\
VFLIP                    & 0.74          & 0.23         & 0.75           & 0.21          & 0.74         & 0.24         & 0.77           & 0.29          & 0.76              & 0.39             \\
\cellcolor{greyL}\alg                     & \cellcolor{greyL}0.86          & \cellcolor{greyL}0.04         & \cellcolor{greyL}0.84           & \cellcolor{greyL}0.07          & \cellcolor{greyL}0.83         & \cellcolor{greyL}0.06         & \cellcolor{greyL}0.84           & \cellcolor{greyL}0.05          & \cellcolor{greyL}0.85              & \cellcolor{greyL}0.14             \\ \hline
\end{tabular}
\label{multi_attacker}
 \vspace{-.05in}
\end{table}

\begin{figure*}
    \centering
    \includegraphics[width=1.0\linewidth]{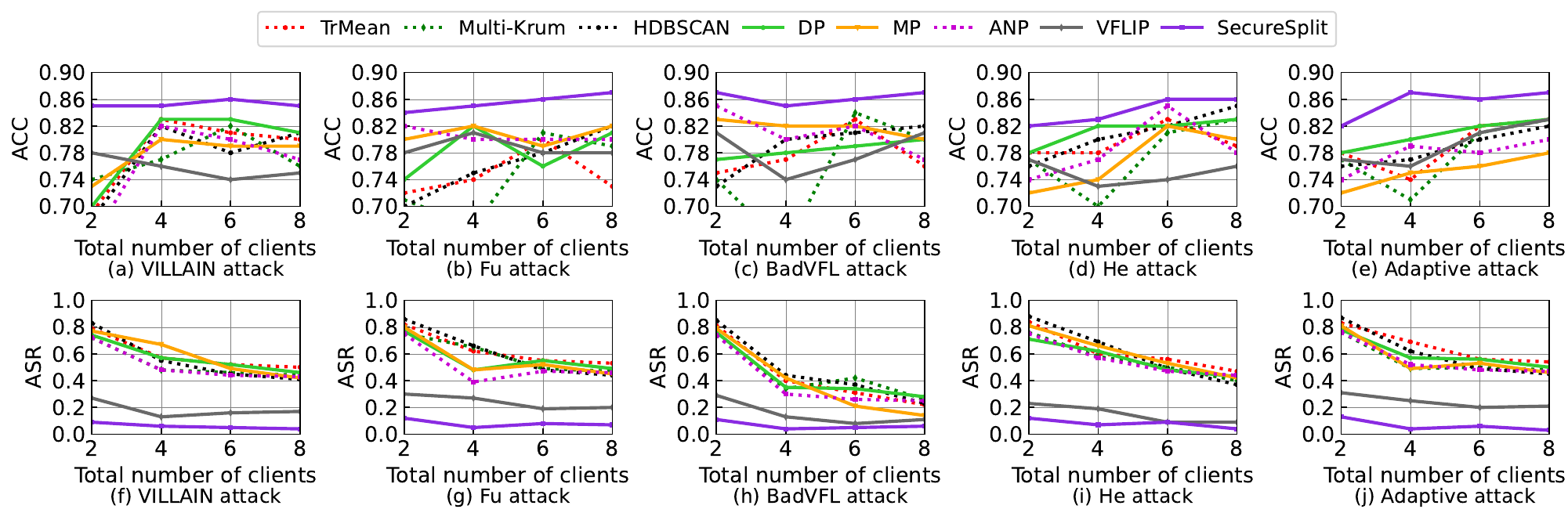} %
    \caption{Impact of the total number of clients, where CIFAR-10 dataset is considered.}
    \label{total_number}
     \vspace{-0.1cm}
\end{figure*}

\myparatight{Impact of different embedding aggregation approaches}%
Table~\ref{aggre} presents the performance of five embedding aggregation methods (embedding aggregation method is the $\mathcal{A}(\cdot)$ in Eq.~(\ref{embed_agg})) under various attack scenarios: Average (computing the element-wise average), Max (selecting the element-wise maximum), Min (choosing the element-wise minimum), Median (picking the element-wise median), and Concatenation (combining embeddings from different clients). The experiments, conducted on the CIFAR-10 dataset, assume identical embedding dimensions across local clients, as per~\cite{bai2023villain}. Under the VILLAIN attack, the Max method yielded the highest ACC (0.87), while the Min method achieved the lowest ASR (0.03). Under the He attack, the Median method performed strongly in both ACC and ASR (0.84 and 0.03, respectively). Overall, \alg offers robust defense across all aggregation methods.

\myparatight{Performance with multiple malicious clients}%
To further assess the robustness of \alg, we conducted an experiment with multiple malicious clients. In this setup, eight clients participate in the training process, three of which are malicious. The experiment is carried out on CIFAR-10. This scenario, involving multiple malicious clients, presents a more challenging and aggressive setup compared to a single malicious client.
Table~\ref{multi_attacker} shows the performance of various defense methods in a scenario with multiple malicious clients, evaluated using ACC and ASR. \alg consistently achieves the best performance, with the highest ACC and lowest ASR across all attacks.

\myparatight{Impact of total number of clients}%
Fig.~\ref{total_number} illustrates the performance of ACC and ASR across various attacks and different total client numbers (2, 4, 6, and 8). As the number of clients increases, the performance of all methods generally improves. In our proposed \alg, the ASR metric consistently performs well, never exceeding 0.10 in any scenario.

\myparatight{Impact of degree of Non-IID}%
By default, we adopt the most extreme Non-IID scenario, where each client holds a completely disjoint set of features. Here, we analyze how varying the degree of feature overlap affects the performance of different defenses under various attacks. Results in Fig.~\ref{image_overlap} in Appendix show that our proposed \alg consistently maintains strong performance across different levels of feature heterogeneity.

\myparatight{Impact of the number of selected clients per round}%
By default, we assume that all clients participate in each training round. In this section, we explore a more practical setting where only a subset of clients joins each round. For example, with four total clients, only one or two may participate in a given round. This introduces the possibility that a malicious client is selected when only one participates. As shown in Fig.~\ref{image_selection} in Appendix, our \alg remains robust under this setting.

\myparatight{Impact of different dimensionality expansion and reduction methods}%
Table~\ref{dimension} presents the performance of various dimensionality reduction and expansion methods in \alg against five attack types on the CIFAR-10 dataset, using ACC and ASR metrics. Dimensionality reduction methods include PCA~\cite{abdi2010principal}, t-SNE~\cite{van2008visualizing}, and factor analysis~\cite{kline2014easy}, while expansion methods include RBF~\cite{scholkopf1997kernel} and feature expansion~\cite{scholkopf2002learning}. 
The results show UMAP performs best among reduction methods,
while PKT excels in expansion methods.

\myparatight{Impact of the poisoning start round}%
Table~\ref{start_posion} in Appendix shows the effect of the starting poison round on the defense method's performance against five different attacks. The results indicate that as the starting poison round increases, ACC values generally improve. For example, under the VILLAIN attack, ACC rises from 0.83 (at round 20) to 0.86 (at rounds 60 and 100).
Overall, our method proves to be a robust defense across all considered scenarios.

\begin{table*}[]
\footnotesize
\caption{Performance of various defenses under different dimensionality expansion and reduction methods. The CIFAR-10 dataset is considered.}
\centering
\begin{tabular}{|cc|cc|cc|cc|cc|cc|}
\hline
\multicolumn{2}{|c|}{\multirow{2}{*}{Dimensionality transformation}}                                                           & \multicolumn{2}{c|}{VILLAIN attack}  & \multicolumn{2}{c|}{Fu attack} & \multicolumn{2}{c|}{BadVFL attack}   & \multicolumn{2}{c|}{He attack} & \multicolumn{2}{c|}{Adaptive attack} \\ \cline{3-12} 
\multicolumn{2}{|c|}{}                                                                                                         & ACC            & ASR           & ACC             & ASR           & ACC            & ASR           & ACC             & ASR           & ACC                & ASR              \\ \hline
\multicolumn{1}{|c|}{\multirow{4}{*}{\begin{tabular}[c]{@{}c@{}}Dimensionality \\ reduction\end{tabular}}} & PCA               & 0.82          & 0.25          & 0.84           & 0.07          & 0.82          & 0.03          & 0.83           & 0.22          & 0.83              & 0.09             \\
\multicolumn{1}{|c|}{}                                                                                     & t-SNE             & 0.84          & 0.19          & 0.83           & 0.05          & 0.86 & 0.02          & 0.85  & 0.27          & 0.85              & 0.05             \\
\multicolumn{1}{|c|}{}                                                                                     & Factor analysis   & 0.83          & 0.17          & 0.81           & 0.08          & 0.83          & 0.06          & 0.82           & 0.08          & 0.82              & 0.05             \\
\multicolumn{1}{|c|}{}                                                                                     & \cellcolor{greyL}UMAP (\alg)              & \cellcolor{greyL}0.85 & \cellcolor{greyL}0.06 & \cellcolor{greyL}0.85  & \cellcolor{greyL}0.05 & \cellcolor{greyL}0.85          & \cellcolor{greyL}0.04 & \cellcolor{greyL}0.83           & \cellcolor{greyL}0.07 & \cellcolor{greyL}0.87     & \cellcolor{greyL}0.08    \\ \hline
\multicolumn{1}{|c|}{Dimensionality}                                                                       & RBF               & 0.82          & 0.12          & 0.83           & 0.07          & 0.84          & 0.04          & 0.84  & 0.13          & 0.82              & 0.13             \\
\multicolumn{1}{|c|}{expansion}                                                                            & Feature expansion & 0.84          & 0.09          & 0.82           & 0.10          & 0.81          & 0.07          & 0.82           & 0.10          & 0.87     & 0.06             \\
\multicolumn{1}{|c|}{}                                                                                     & \cellcolor{greyL}PKT (\alg)               & \cellcolor{greyL}0.85 & \cellcolor{greyL}0.06 & \cellcolor{greyL}0.85  & \cellcolor{greyL}0.05 & \cellcolor{greyL}0.85 & \cellcolor{greyL}0.04 & \cellcolor{greyL}0.83           & \cellcolor{greyL}0.07 & \cellcolor{greyL}0.87     & \cellcolor{greyL}0.08    \\ \hline
\end{tabular}
\label{dimension}
 \vspace{-.05in}
\end{table*}

\myparatight{Different variants of \alg}%
To assess the effectiveness of each component, we created seven distinct variants. Specifically, Variant I uses $R$ for filtering on embedding $\bm{E}$ with Eq.~(\ref{compute_R}), selecting half of the embedding samples closest to the median point of $\bm{E}$. Variant II uses $R_{\text{opt}}$ for filtering on embedding $\bm{E}$ with Eq.~(\ref{our_filter}), where we compute the radius $R_{\text{opt}}$ for the embedding set $E$ and select samples within this radius. Variant III applies filtering based on $R$ on embedding $\bm{U}$, while Variant IV uses $R_{\text{opt}}$ for filtering on embedding $\bm{U}$. Variant V filters based on $R$ on embedding $\bm{H}$, and Variant VI uses K-means for filtering on embedding $\bm{H}$. These variants allow a systematic evaluation of how dimensionality reduction, expansion, and aggregation strategies affect model performance. 
Table~\ref{various_components} in Appendix clearly shows that each component is essential, and only their combination ensures robust defense.

\myparatight{Computational overhead of different methods}%
Figs.~\ref{time_cifar10}-\ref{time_cinic10} in the Appendix present the runtime comparison of various methods across four datasets under the He attack. Computational overhead is quantified by the total execution time of each method. The results show that \alg introduces only minimal additional cost relative to the baseline setting without any defense.

\myparatight{Compare \alg with SafeSplit~\cite{rieger2025safesplit}}%
We do not compare \alg with SafeSplit~\cite{rieger2025safesplit} in the above experiments, as SafeSplit is specifically designed for U-shaped SL with sequential client training and cannot be applied to the standard SL setting, where clients train in parallel. See Appendix~\ref{app_USL} for more details on U-shaped SL. To fairly assess \alg, we adapt it and other standard SL methods to the U-shaped SL setup, following the configuration in~\cite{rieger2025safesplit} with 10 clients (2 malicious) and their backdoor attack. As shown in Table~\ref{U_results} in Appendix, \alg generalizes well to this setting and outperforms SafeSplit.


\section{Conclusion}
\label{sec:conclusion}

We propose \alg, a new defense against backdoor attacks in SL. \alg enhances detection by reshaping embeddings to highlight differences between benign and malicious inputs. It then applies a dynamic majority-based filter to identify and remove compromised embeddings. Comprehensive experiments confirm the effectiveness of \alg.

\begin{acks}
We thank the anonymous reviewers for their comments.
\end{acks}

\bibliographystyle{ACM-Reference-Format}
\balance
\bibliography{refs}


\begin{thebibliography}{49}


\ifx \showCODEN    \undefined \def \showCODEN     #1{\unskip}     \fi
\ifx \showISBNx    \undefined \def \showISBNx     #1{\unskip}     \fi
\ifx \showISBNxiii \undefined \def \showISBNxiii  #1{\unskip}     \fi
\ifx \showISSN     \undefined \def \showISSN      #1{\unskip}     \fi
\ifx \showLCCN     \undefined \def \showLCCN      #1{\unskip}     \fi
\ifx \shownote     \undefined \def \shownote      #1{#1}          \fi
\ifx \showarticletitle \undefined \def \showarticletitle #1{#1}   \fi
\ifx \showURL      \undefined \def \showURL       {\relax}        \fi
\providecommand\bibfield[2]{#2}
\providecommand\bibinfo[2]{#2}
\providecommand\natexlab[1]{#1}
\providecommand\showeprint[2][]{arXiv:#2}

\bibitem[web({[n.\,d.]})]%
        {webank}
 \bibinfo{year}{[n.\,d.]}\natexlab{}.
\newblock \bibinfo{booktitle}{\emph{Utilization of FATE in Risk Management of
  Credit in Small and Micro Enterprises}}.
\newblock
\urldef\tempurl%
\url{https://www.fedai.org/cases/utilization-of-fate-in-risk-management-of-credit-in-small-and-micro-\\enterprises/}
\showURL{%
\tempurl}


\bibitem[Abadi et~al\mbox{.}(2016)]%
        {abadi2016deep}
\bibfield{author}{\bibinfo{person}{Martin Abadi}, \bibinfo{person}{Andy Chu},
  \bibinfo{person}{Ian Goodfellow}, \bibinfo{person}{H~Brendan McMahan},
  \bibinfo{person}{Ilya Mironov}, \bibinfo{person}{Kunal Talwar}, {and}
  \bibinfo{person}{Li Zhang}.} \bibinfo{year}{2016}\natexlab{}.
\newblock \showarticletitle{Deep learning with differential privacy}. In
  \bibinfo{booktitle}{\emph{CCS}}.
\newblock


\bibitem[Abdi and Williams(2010)]%
        {abdi2010principal}
\bibfield{author}{\bibinfo{person}{Herv{\'e} Abdi} {and}
  \bibinfo{person}{Lynne~J Williams}.} \bibinfo{year}{2010}\natexlab{}.
\newblock \showarticletitle{Principal component analysis}.
\newblock \bibinfo{journal}{\emph{Wiley interdisciplinary reviews:
  computational statistics}} \bibinfo{volume}{2}, \bibinfo{number}{4}
  (\bibinfo{year}{2010}), \bibinfo{pages}{433--459}.
\newblock


\bibitem[Bai et~al\mbox{.}(2023)]%
        {bai2023villain}
\bibfield{author}{\bibinfo{person}{Yijie Bai}, \bibinfo{person}{Yanjiao Chen},
  \bibinfo{person}{Hanlei Zhang}, \bibinfo{person}{Wenyuan Xu},
  \bibinfo{person}{Haiqin Weng}, {and} \bibinfo{person}{Dou Goodman}.}
  \bibinfo{year}{2023}\natexlab{}.
\newblock \showarticletitle{VILLAIN: Backdoor attacks against vertical split
  learning}. In \bibinfo{booktitle}{\emph{USENIX Security Symposium}}.
\newblock


\bibitem[Blanchard et~al\mbox{.}(2017)]%
        {blanchard2017machine}
\bibfield{author}{\bibinfo{person}{Peva Blanchard}, \bibinfo{person}{El~Mahdi
  El~Mhamdi}, \bibinfo{person}{Rachid Guerraoui}, {and} \bibinfo{person}{Julien
  Stainer}.} \bibinfo{year}{2017}\natexlab{}.
\newblock \showarticletitle{Machine learning with adversaries: Byzantine
  tolerant gradient descent}. In \bibinfo{booktitle}{\emph{NeurIPS}}.
\newblock


\bibitem[Cao et~al\mbox{.}(2021)]%
        {cao2020fltrust}
\bibfield{author}{\bibinfo{person}{Xiaoyu Cao}, \bibinfo{person}{Minghong
  Fang}, \bibinfo{person}{Jia Liu}, {and} \bibinfo{person}{Neil~Zhenqiang
  Gong}.} \bibinfo{year}{2021}\natexlab{}.
\newblock \showarticletitle{Fltrust: Byzantine-robust federated learning via
  trust bootstrapping}. In \bibinfo{booktitle}{\emph{NDSS}}.
\newblock


\bibitem[Chen et~al\mbox{.}(2024)]%
        {chen2024contributions}
\bibfield{author}{\bibinfo{person}{Yiwei Chen}, \bibinfo{person}{Kaiyu Li},
  \bibinfo{person}{Guoliang Li}, {and} \bibinfo{person}{Yong Wang}.}
  \bibinfo{year}{2024}\natexlab{}.
\newblock \showarticletitle{Contributions Estimation in Federated Learning: A
  Comprehensive Experimental Evaluation}. In \bibinfo{booktitle}{\emph{VLDB}}.
\newblock


\bibitem[Cho et~al\mbox{.}(2024)]%
        {cho2024vflip}
\bibfield{author}{\bibinfo{person}{Yungi Cho}, \bibinfo{person}{Woorim Han},
  \bibinfo{person}{Miseon Yu}, \bibinfo{person}{Younghan Lee},
  \bibinfo{person}{Ho Bae}, {and} \bibinfo{person}{Yunheung Paek}.}
  \bibinfo{year}{2024}\natexlab{}.
\newblock \showarticletitle{VFLIP: A Backdoor Defense for Vertical Federated
  Learning via Identification and Purification}. In
  \bibinfo{booktitle}{\emph{European Symposium on Research in Computer
  Security}}.
\newblock


\bibitem[Darlow et~al\mbox{.}(2018)]%
        {darlow2018cinic}
\bibfield{author}{\bibinfo{person}{Luke~N Darlow}, \bibinfo{person}{Elliot~J
  Crowley}, \bibinfo{person}{Antreas Antoniou}, {and} \bibinfo{person}{Amos~J
  Storkey}.} \bibinfo{year}{2018}\natexlab{}.
\newblock \showarticletitle{Cinic-10 is not imagenet or cifar-10}.
\newblock \bibinfo{journal}{\emph{arXiv preprint arXiv:1810.03505}}
  (\bibinfo{year}{2018}).
\newblock


\bibitem[Dou et~al\mbox{.}(2025)]%
        {dou2025toward}
\bibfield{author}{\bibinfo{person}{Zhihao Dou}, \bibinfo{person}{Jiaqi Wang},
  \bibinfo{person}{Wei Sun}, \bibinfo{person}{Zhuqing Liu}, {and}
  \bibinfo{person}{Minghong Fang}.} \bibinfo{year}{2025}\natexlab{}.
\newblock \showarticletitle{Toward Malicious Clients Detection in Federated
  Learning}. In \bibinfo{booktitle}{\emph{ASIACCS}}.
\newblock


\bibitem[Fang et~al\mbox{.}(2020)]%
        {fang2020local}
\bibfield{author}{\bibinfo{person}{Minghong Fang}, \bibinfo{person}{Xiaoyu
  Cao}, \bibinfo{person}{Jinyuan Jia}, {and} \bibinfo{person}{Neil Gong}.}
  \bibinfo{year}{2020}\natexlab{}.
\newblock \showarticletitle{Local model poisoning attacks to Byzantine-robust
  federated learning}. In \bibinfo{booktitle}{\emph{USENIX Security
  Symposium}}.
\newblock


\bibitem[Fang et~al\mbox{.}(2022)]%
        {fang2022aflguard}
\bibfield{author}{\bibinfo{person}{Minghong Fang}, \bibinfo{person}{Jia Liu},
  \bibinfo{person}{Neil~Zhenqiang Gong}, {and} \bibinfo{person}{Elizabeth~S
  Bentley}.} \bibinfo{year}{2022}\natexlab{}.
\newblock \showarticletitle{Aflguard: Byzantine-robust asynchronous federated
  learning}. In \bibinfo{booktitle}{\emph{ACSAC}}.
\newblock


\bibitem[Fang et~al\mbox{.}(2025a)]%
        {fang2025byzantine}
\bibfield{author}{\bibinfo{person}{Minghong Fang}, \bibinfo{person}{Zhuqing
  Liu}, \bibinfo{person}{Xuecen Zhao}, {and} \bibinfo{person}{Jia Liu}.}
  \bibinfo{year}{2025}\natexlab{a}.
\newblock \showarticletitle{Byzantine-Robust Federated Learning over
  Ring-All-Reduce Distributed Computing}. In \bibinfo{booktitle}{\emph{The Web
  Conference}}.
\newblock


\bibitem[Fang et~al\mbox{.}(2025b)]%
        {fang2025we}
\bibfield{author}{\bibinfo{person}{Minghong Fang}, \bibinfo{person}{Seyedsina
  Nabavirazavi}, \bibinfo{person}{Zhuqing Liu}, \bibinfo{person}{Wei Sun},
  \bibinfo{person}{Sundararaja~Sitharama Iyengar}, {and} \bibinfo{person}{Haibo
  Yang}.} \bibinfo{year}{2025}\natexlab{b}.
\newblock \showarticletitle{Do we really need to design new byzantine-robust
  aggregation rules?}. In \bibinfo{booktitle}{\emph{NDSS}}.
\newblock


\bibitem[Fang et~al\mbox{.}(2025c)]%
        {fang2025provably}
\bibfield{author}{\bibinfo{person}{Minghong Fang}, \bibinfo{person}{Xilong
  Wang}, {and} \bibinfo{person}{Neil~Zhenqiang Gong}.}
  \bibinfo{year}{2025}\natexlab{c}.
\newblock \showarticletitle{Provably Robust Federated Reinforcement Learning}.
  In \bibinfo{booktitle}{\emph{The Web Conference}}.
\newblock


\bibitem[Fang et~al\mbox{.}(2024)]%
        {fang2024byzantine}
\bibfield{author}{\bibinfo{person}{Minghong Fang}, \bibinfo{person}{Zifan
  Zhang}, \bibinfo{person}{Prashant Khanduri}, \bibinfo{person}{Jia Liu},
  \bibinfo{person}{Songtao Lu}, \bibinfo{person}{Yuchen Liu},
  \bibinfo{person}{Neil Gong}, {et~al\mbox{.}}}
  \bibinfo{year}{2024}\natexlab{}.
\newblock \showarticletitle{Byzantine-robust decentralized federated learning}.
  In \bibinfo{booktitle}{\emph{CCS}}.
\newblock


\bibitem[Fu et~al\mbox{.}(2022b)]%
        {fu2022label}
\bibfield{author}{\bibinfo{person}{Chong Fu}, \bibinfo{person}{Xuhong Zhang},
  \bibinfo{person}{Shouling Ji}, \bibinfo{person}{Jinyin Chen},
  \bibinfo{person}{Jingzheng Wu}, \bibinfo{person}{Shanqing Guo},
  \bibinfo{person}{Jun Zhou}, \bibinfo{person}{Alex~X Liu}, {and}
  \bibinfo{person}{Ting Wang}.} \bibinfo{year}{2022}\natexlab{b}.
\newblock \showarticletitle{Label inference attacks against vertical federated
  learning}. In \bibinfo{booktitle}{\emph{USENIX Security Symposium}}.
\newblock


\bibitem[Fu et~al\mbox{.}(2022a)]%
        {fu2022blindfl}
\bibfield{author}{\bibinfo{person}{Fangcheng Fu}, \bibinfo{person}{Huanran
  Xue}, \bibinfo{person}{Yong Cheng}, \bibinfo{person}{Yangyu Tao}, {and}
  \bibinfo{person}{Bin Cui}.} \bibinfo{year}{2022}\natexlab{a}.
\newblock \showarticletitle{Blindfl: Vertical federated machine learning
  without peeking into your data}. In \bibinfo{booktitle}{\emph{SIGMOD}}.
\newblock


\bibitem[He et~al\mbox{.}(2023)]%
        {he2023backdoor}
\bibfield{author}{\bibinfo{person}{Ying He}, \bibinfo{person}{Zhili Shen},
  \bibinfo{person}{Jingyu Hua}, \bibinfo{person}{Qixuan Dong},
  \bibinfo{person}{Jiacheng Niu}, \bibinfo{person}{Wei Tong},
  \bibinfo{person}{Xu Huang}, \bibinfo{person}{Chen Li}, {and}
  \bibinfo{person}{Sheng Zhong}.} \bibinfo{year}{2023}\natexlab{}.
\newblock \showarticletitle{Backdoor attack against split neural network-based
  vertical federated learning}. In \bibinfo{booktitle}{\emph{IEEE Transactions
  on Information Forensics and Security}}.
\newblock


\bibitem[Howard and Gugger(2020)]%
        {howard2020fastai}
\bibfield{author}{\bibinfo{person}{Jeremy Howard} {and}
  \bibinfo{person}{Sylvain Gugger}.} \bibinfo{year}{2020}\natexlab{}.
\newblock \showarticletitle{Fastai: a layered API for deep learning}.
\newblock \bibinfo{journal}{\emph{Information}} \bibinfo{volume}{11},
  \bibinfo{number}{2} (\bibinfo{year}{2020}), \bibinfo{pages}{108}.
\newblock


\bibitem[Kline(2014)]%
        {kline2014easy}
\bibfield{author}{\bibinfo{person}{Paul Kline}.}
  \bibinfo{year}{2014}\natexlab{}.
\newblock \bibinfo{booktitle}{\emph{An easy guide to factor analysis}}.
\newblock \bibinfo{publisher}{Routledge}.
\newblock


\bibitem[Krizhevsky and Hinton(2009)]%
        {cifar10data}
\bibfield{author}{\bibinfo{person}{A. Krizhevsky} {and} \bibinfo{person}{G.
  Hinton}.} \bibinfo{year}{2009}\natexlab{}.
\newblock \showarticletitle{Learning multiple layers of features from tiny
  images}.
\newblock \bibinfo{journal}{\emph{Handbook of Systemic Autoimmune Diseases}}
  (\bibinfo{year}{2009}).
\newblock


\bibitem[LeCun et~al\mbox{.}(1998a)]%
        {lecun1998gradient}
\bibfield{author}{\bibinfo{person}{Yann LeCun}, \bibinfo{person}{L{\'e}on
  Bottou}, \bibinfo{person}{Yoshua Bengio}, {and} \bibinfo{person}{Patrick
  Haffner}.} \bibinfo{year}{1998}\natexlab{a}.
\newblock \showarticletitle{Gradient-based learning applied to document
  recognition}.
\newblock \bibinfo{journal}{\emph{Proc. IEEE}} \bibinfo{volume}{86},
  \bibinfo{number}{11} (\bibinfo{year}{1998}), \bibinfo{pages}{2278--2324}.
\newblock


\bibitem[LeCun et~al\mbox{.}(1998b)]%
        {mnist}
\bibfield{author}{\bibinfo{person}{Yann LeCun}, \bibinfo{person}{Corinna
  Cortes}, {and} \bibinfo{person}{CJ Burges}.}
  \bibinfo{year}{1998}\natexlab{b}.
\newblock \showarticletitle{MNIST handwritten digit database}.
\newblock \bibinfo{journal}{\emph{Available: http://yann. lecun.
  com/exdb/mnist}} (\bibinfo{year}{1998}).
\newblock


\bibitem[Liu et~al\mbox{.}(2018)]%
        {liu2018fine}
\bibfield{author}{\bibinfo{person}{Kang Liu}, \bibinfo{person}{Brendan
  Dolan-Gavitt}, {and} \bibinfo{person}{Siddharth Garg}.}
  \bibinfo{year}{2018}\natexlab{}.
\newblock \showarticletitle{Fine-pruning: Defending against backdooring attacks
  on deep neural networks}. In \bibinfo{booktitle}{\emph{RAID}}.
\newblock


\bibitem[Liu et~al\mbox{.}(2020)]%
        {liu2020asymmetrical}
\bibfield{author}{\bibinfo{person}{Yang Liu}, \bibinfo{person}{Xiong Zhang},
  {and} \bibinfo{person}{Libin Wang}.} \bibinfo{year}{2020}\natexlab{}.
\newblock \showarticletitle{Asymmetrical vertical federated learning}.
\newblock \bibinfo{journal}{\emph{arXiv preprint arXiv:2004.07427}}
  (\bibinfo{year}{2020}).
\newblock


\bibitem[Malzer and Baum(2020)]%
        {malzer2020hybrid}
\bibfield{author}{\bibinfo{person}{Claudia Malzer} {and}
  \bibinfo{person}{Marcus Baum}.} \bibinfo{year}{2020}\natexlab{}.
\newblock \showarticletitle{A hybrid approach to hierarchical density-based
  cluster selection}. In \bibinfo{booktitle}{\emph{MFI}}.
\newblock


\bibitem[McInnes et~al\mbox{.}(2017)]%
        {mcinnes2017hdbscan}
\bibfield{author}{\bibinfo{person}{Leland McInnes}, \bibinfo{person}{John
  Healy}, \bibinfo{person}{Steve Astels}, {et~al\mbox{.}}}
  \bibinfo{year}{2017}\natexlab{}.
\newblock \showarticletitle{hdbscan: Hierarchical density based clustering.}
\newblock \bibinfo{journal}{\emph{J. Open Source Softw.}} \bibinfo{volume}{2},
  \bibinfo{number}{11} (\bibinfo{year}{2017}), \bibinfo{pages}{205}.
\newblock


\bibitem[McInnes et~al\mbox{.}(2018)]%
        {mcinnes2018umap}
\bibfield{author}{\bibinfo{person}{Leland McInnes}, \bibinfo{person}{John
  Healy}, {and} \bibinfo{person}{James Melville}.}
  \bibinfo{year}{2018}\natexlab{}.
\newblock \showarticletitle{Umap: Uniform manifold approximation and projection
  for dimension reduction}.
\newblock \bibinfo{journal}{\emph{arXiv preprint arXiv:1802.03426}}.
\newblock


\bibitem[McMahan et~al\mbox{.}(2017)]%
        {mcmahan2017communication}
\bibfield{author}{\bibinfo{person}{H.~Brendan McMahan}, \bibinfo{person}{Eider
  Moore}, \bibinfo{person}{Daniel Ramage}, \bibinfo{person}{Seth Hampson},
  {and} \bibinfo{person}{Blaise~Ag{\"u}era y Arcas}.}
  \bibinfo{year}{2017}\natexlab{}.
\newblock \showarticletitle{Communication-Efficient Learning of Deep Networks
  from Decentralized Data}. In \bibinfo{booktitle}{\emph{AISTATS}}.
\newblock


\bibitem[Mo et~al\mbox{.}(2025)]%
        {mo2025find}
\bibfield{author}{\bibinfo{person}{Wenjin Mo}, \bibinfo{person}{Zhiyuan Li},
  \bibinfo{person}{Minghong Fang}, {and} \bibinfo{person}{Mingwei Fang}.}
  \bibinfo{year}{2025}\natexlab{}.
\newblock \showarticletitle{Find a Scapegoat: Poisoning Membership Inference
  Attack and Defense to Federated Learning}. In
  \bibinfo{booktitle}{\emph{ICCV}}.
\newblock


\bibitem[Naseri et~al\mbox{.}(2024)]%
        {naseri2024badvfl}
\bibfield{author}{\bibinfo{person}{Mohammad Naseri}, \bibinfo{person}{Yufei
  Han}, {and} \bibinfo{person}{Emiliano De~Cristofaro}.}
  \bibinfo{year}{2024}\natexlab{}.
\newblock \showarticletitle{Badvfl: Backdoor attacks in vertical federated
  learning}. In \bibinfo{booktitle}{\emph{IEEE Symposium on Security and
  Privacy}}.
\newblock


\bibitem[Paulik et~al\mbox{.}(2021)]%
        {paulik2021federated}
\bibfield{author}{\bibinfo{person}{Matthias Paulik}, \bibinfo{person}{Matt
  Seigel}, \bibinfo{person}{Henry Mason}, \bibinfo{person}{Dominic Telaar},
  \bibinfo{person}{Joris Kluivers}, \bibinfo{person}{Rogier van Dalen},
  \bibinfo{person}{Chi~Wai Lau}, \bibinfo{person}{Luke Carlson},
  \bibinfo{person}{Filip Granqvist}, \bibinfo{person}{Chris Vandevelde},
  {et~al\mbox{.}}} \bibinfo{year}{2021}\natexlab{}.
\newblock \showarticletitle{Federated evaluation and tuning for on-device
  personalization: System design \& applications}.
\newblock \bibinfo{journal}{\emph{arXiv preprint arXiv:2102.08503}}
  (\bibinfo{year}{2021}).
\newblock


\bibitem[Poirot et~al\mbox{.}(2019)]%
        {poirot2019split}
\bibfield{author}{\bibinfo{person}{Maarten~G Poirot}, \bibinfo{person}{Praneeth
  Vepakomma}, \bibinfo{person}{Ken Chang}, \bibinfo{person}{Jayashree
  Kalpathy-Cramer}, \bibinfo{person}{Rajiv Gupta}, {and}
  \bibinfo{person}{Ramesh Raskar}.} \bibinfo{year}{2019}\natexlab{}.
\newblock \showarticletitle{Split learning for collaborative deep learning in
  healthcare}.
\newblock \bibinfo{journal}{\emph{arXiv preprint arXiv:1912.12115}}
  (\bibinfo{year}{2019}).
\newblock


\bibitem[Rieger et~al\mbox{.}(2025)]%
        {rieger2025safesplit}
\bibfield{author}{\bibinfo{person}{Phillip Rieger}, \bibinfo{person}{Alessandro
  Pegoraro}, \bibinfo{person}{Kavita Kumari}, \bibinfo{person}{Tigist Abera},
  \bibinfo{person}{Jonathan Knauer}, {and} \bibinfo{person}{Ahmad-Reza
  Sadeghi}.} \bibinfo{year}{2025}\natexlab{}.
\newblock \showarticletitle{SafeSplit: A Novel Defense Against Client-Side
  Backdoor Attacks in Split Learning}. In \bibinfo{booktitle}{\emph{NDSS}}.
\newblock


\bibitem[Romanini et~al\mbox{.}(2021)]%
        {romanini2021pyvertical}
\bibfield{author}{\bibinfo{person}{Daniele Romanini},
  \bibinfo{person}{Adam~James Hall}, \bibinfo{person}{Pavlos Papadopoulos},
  \bibinfo{person}{Tom Titcombe}, \bibinfo{person}{Abbas Ismail},
  \bibinfo{person}{Tudor Cebere}, \bibinfo{person}{Robert Sandmann},
  \bibinfo{person}{Robin Roehm}, {and} \bibinfo{person}{Michael~A Hoeh}.}
  \bibinfo{year}{2021}\natexlab{}.
\newblock \showarticletitle{Pyvertical: A vertical federated learning framework
  for multi-headed splitnn}.
\newblock \bibinfo{journal}{\emph{arXiv preprint arXiv:2104.00489}}
  (\bibinfo{year}{2021}).
\newblock


\bibitem[Sch{\"o}lkopf et~al\mbox{.}(1997)]%
        {scholkopf1997kernel}
\bibfield{author}{\bibinfo{person}{Bernhard Sch{\"o}lkopf},
  \bibinfo{person}{Alexander Smola}, {and} \bibinfo{person}{Klaus-Robert
  M{\"u}ller}.} \bibinfo{year}{1997}\natexlab{}.
\newblock \showarticletitle{Kernel principal component analysis}. In
  \bibinfo{booktitle}{\emph{ICANN}}.
\newblock


\bibitem[Sch{\"o}lkopf and Smola(2002)]%
        {scholkopf2002learning}
\bibfield{author}{\bibinfo{person}{Bernhard Sch{\"o}lkopf} {and}
  \bibinfo{person}{Alexander~J Smola}.} \bibinfo{year}{2002}\natexlab{}.
\newblock \bibinfo{booktitle}{\emph{Learning with kernels: support vector
  machines, regularization, optimization, and beyond}}.
\newblock \bibinfo{publisher}{MIT press}.
\newblock


\bibitem[Singh et~al\mbox{.}(2019)]%
        {singh2019detailed}
\bibfield{author}{\bibinfo{person}{Abhishek Singh}, \bibinfo{person}{Praneeth
  Vepakomma}, \bibinfo{person}{Otkrist Gupta}, {and} \bibinfo{person}{Ramesh
  Raskar}.} \bibinfo{year}{2019}\natexlab{}.
\newblock \showarticletitle{Detailed comparison of communication efficiency of
  split learning and federated learning}.
\newblock \bibinfo{journal}{\emph{arXiv preprint arXiv:1909.09145}}
  (\bibinfo{year}{2019}).
\newblock


\bibitem[Thapa et~al\mbox{.}(2022)]%
        {thapa2022splitfed}
\bibfield{author}{\bibinfo{person}{Chandra Thapa}, \bibinfo{person}{Pathum
  Chamikara~Mahawaga Arachchige}, \bibinfo{person}{Seyit Camtepe}, {and}
  \bibinfo{person}{Lichao Sun}.} \bibinfo{year}{2022}\natexlab{}.
\newblock \showarticletitle{Splitfed: When federated learning meets split
  learning}. In \bibinfo{booktitle}{\emph{AAAI}}.
\newblock


\bibitem[Van~der Maaten and Hinton(2008)]%
        {van2008visualizing}
\bibfield{author}{\bibinfo{person}{Laurens Van~der Maaten} {and}
  \bibinfo{person}{Geoffrey Hinton}.} \bibinfo{year}{2008}\natexlab{}.
\newblock \showarticletitle{Visualizing data using t-SNE.}
\newblock \bibinfo{journal}{\emph{Journal of machine learning research}}
  \bibinfo{volume}{9}, \bibinfo{number}{11} (\bibinfo{year}{2008}).
\newblock


\bibitem[Vepakomma et~al\mbox{.}(2018)]%
        {vepakomma2018split}
\bibfield{author}{\bibinfo{person}{Praneeth Vepakomma},
  \bibinfo{person}{Otkrist Gupta}, \bibinfo{person}{Tristan Swedish}, {and}
  \bibinfo{person}{Ramesh Raskar}.} \bibinfo{year}{2018}\natexlab{}.
\newblock \showarticletitle{Split learning for health: Distributed deep
  learning without sharing raw patient data}.
\newblock \bibinfo{journal}{\emph{arXiv preprint arXiv:1812.00564}}
  (\bibinfo{year}{2018}).
\newblock


\bibitem[Wang et~al\mbox{.}(2025)]%
        {wang2025poisoning}
\bibfield{author}{\bibinfo{person}{Wenbin Wang}, \bibinfo{person}{Qiwen Ma},
  \bibinfo{person}{Zifan Zhang}, \bibinfo{person}{Yuchen Liu},
  \bibinfo{person}{Zhuqing Liu}, {and} \bibinfo{person}{Minghong Fang}.}
  \bibinfo{year}{2025}\natexlab{}.
\newblock \showarticletitle{Poisoning attacks and defenses to federated
  unlearning}. In \bibinfo{booktitle}{\emph{The Web Conference}}.
\newblock


\bibitem[Wei{\ss}e et~al\mbox{.}(2006)]%
        {weisse2006kernel}
\bibfield{author}{\bibinfo{person}{Alexander Wei{\ss}e},
  \bibinfo{person}{Gerhard Wellein}, \bibinfo{person}{Andreas Alvermann}, {and}
  \bibinfo{person}{Holger Fehske}.} \bibinfo{year}{2006}\natexlab{}.
\newblock \showarticletitle{The kernel polynomial method}. In
  \bibinfo{booktitle}{\emph{Reviews of modern physics}}.
\newblock


\bibitem[Wu and Wang(2021)]%
        {wu2021adversarial}
\bibfield{author}{\bibinfo{person}{Dongxian Wu} {and} \bibinfo{person}{Yisen
  Wang}.} \bibinfo{year}{2021}\natexlab{}.
\newblock \showarticletitle{Adversarial neuron pruning purifies backdoored deep
  models}. In \bibinfo{booktitle}{\emph{NeurIPS}}.
\newblock


\bibitem[Xie et~al\mbox{.}(2024)]%
        {xie2024fedredefense}
\bibfield{author}{\bibinfo{person}{Yueqi Xie}, \bibinfo{person}{Minghong Fang},
  {and} \bibinfo{person}{Neil~Zhenqiang Gong}.}
  \bibinfo{year}{2024}\natexlab{}.
\newblock \showarticletitle{Fedredefense: Defending against model poisoning
  attacks for federated learning using model update reconstruction error}.
  ICML.
\newblock


\bibitem[Xie et~al\mbox{.}(2022)]%
        {xie2022federatedscope}
\bibfield{author}{\bibinfo{person}{Yuexiang Xie}, \bibinfo{person}{Zhen Wang},
  \bibinfo{person}{Dawei Gao}, \bibinfo{person}{Daoyuan Chen},
  \bibinfo{person}{Liuyi Yao}, \bibinfo{person}{Weirui Kuang},
  \bibinfo{person}{Yaliang Li}, \bibinfo{person}{Bolin Ding}, {and}
  \bibinfo{person}{Jingren Zhou}.} \bibinfo{year}{2022}\natexlab{}.
\newblock \showarticletitle{Federatedscope: A flexible federated learning
  platform for heterogeneity}. In \bibinfo{booktitle}{\emph{VLDB}}.
\newblock


\bibitem[Yin et~al\mbox{.}(2018)]%
        {yin2018byzantine}
\bibfield{author}{\bibinfo{person}{Dong Yin}, \bibinfo{person}{Yudong Chen},
  \bibinfo{person}{Ramchandran Kannan}, {and} \bibinfo{person}{Peter
  Bartlett}.} \bibinfo{year}{2018}\natexlab{}.
\newblock \showarticletitle{Byzantine-robust distributed learning: Towards
  optimal statistical rates}. In \bibinfo{booktitle}{\emph{ICML}}.
\newblock


\bibitem[Zhang et~al\mbox{.}(2024)]%
        {zhang2024securing}
\bibfield{author}{\bibinfo{person}{Zifan Zhang}, \bibinfo{person}{Minghong
  Fang}, \bibinfo{person}{Mingzhe Chen}, \bibinfo{person}{Gaolei Li},
  \bibinfo{person}{Xi Lin}, {and} \bibinfo{person}{Yuchen Liu}.}
  \bibinfo{year}{2024}\natexlab{}.
\newblock \showarticletitle{Securing distributed network digital twin systems
  against model poisoning attacks}. In \bibinfo{booktitle}{\emph{IEEE Internet
  of Things Journal}}.
\newblock


\end{thebibliography}


\appendix

\begin{algorithm}[t]
\caption{Adaptive attack.}
\label{alg:adaptive_attack}
\textbf{Input:} Poisoned embedding $\bar{\bm{E}}_k$, initial trigger magnitude \(\lambda_0\), median $\bm{\Lambda}$ of $\bm{H}$, adaptive radius $R_{\text{adp}}$, and  tolerance threshold $\epsilon_{\text{tol}}$.  \\
\textbf{Output:} Optimized trigger magnitude \(\lambda_{\text{opt}}\).
\begin{algorithmic}[1]
\State Initialize \(\lambda_{\text{max}} \gets \lambda_0\), \(\lambda_{\text{min}} \gets 0\), \(\lambda_{\text{opt}} \gets 0\)
\While{$\lambda_{\text{max}} - \lambda_{\text{min}} > \epsilon_{\text{tol}}$}  \Comment{Tolerance threshold for convergence}
    \State $\lambda \gets (\lambda_{\text{max}} + \lambda_{\text{min}}) / 2$  \Comment{Binary search on trigger magnitude}
    \State Compute updated poisoned embedding $\bar{\bm{E}}_k$ with $\lambda$  
    \If{$\| \bar{\bm{E}}_k - \bm{\Lambda} \|_2 \leq R_{\text{adp}}$}  \Comment{Check defense condition}
        \State $\lambda_{\text{opt}} \gets \lambda$  \Comment{Update optimal trigger magnitude}
        \State $\lambda_{\text{min}} \gets \lambda$  \Comment{Increase lower bound}
    \Else
        \State $\lambda_{\text{max}} \gets \lambda$  \Comment{Decrease upper bound}
    \EndIf
\EndWhile
\State \textbf{return} $\lambda_{\text{opt}}$
\end{algorithmic}
\end{algorithm}

\section{Differences between Horizontal Federated Learning (HFL), Vertical Federated Learning (VFL), and Split Learning (SL)}
\label{app_fl_diff}

\myparatight{a)~Horizontal federated learning (HFL)}%
In HFL, the dataset is divided by samples among various clients, where each client possesses records corresponding to different individuals but with an identical set of features. For example, in a dataset concerning customers, separate entities such as hospitals or banks might store information for different patients or clients, yet each record contains the same attributes such as age, gender, and income.

\myparatight{b)~Vertical federated learning (VFL)}%
VFL is applicable when data is split by features rather than by samples. In this setting, each participating client owns a distinct set of attributes for the same group of individuals. For example, consider two organizations collaborating on shared customer data: one institution may store personal details such as name, age, and gender, while another retains financial information like purchase records and payment history. Although the attributes differ across organizations, they correspond to the same underlying user base.

\myparatight{c)~Split learning (SL)}%
In SL, data is also partitioned by features rather than by samples. In this setting, each participating client holds a distinct subset of features for the same set of samples. Unlike traditional VFL, SL involves a single model that is divided between the client and the server. For example, a deep learning model may be split such that the client computes the bottom model, while the server processes the top model.

\section{Details of Datasets}
\label{app_dataset}

\myparatight{a)~CIFAR-10~\cite{cifar10data}}%
It contains 60,000 color images in 10 classes, with 50,000 for training and 10,000 for testing.

\myparatight{b)~ImageNette~\cite{howard2020fastai}}%
ImageNette is a 10-class subset of the ImageNet dataset, consisting of 9,469 training images and 3,925 testing images.

\myparatight{c)~MNIST~\cite{mnist}}%
~MNIST is a handwritten digit dataset with 10 classes, including 60,000 training and 10,000 testing images.

\myparatight{d)~CINIC-10~\cite{darlow2018cinic}}%
CINIC-10 includes 270,000 color images in 10 classes, designed as an extension of CIFAR-10.

\section{Details of Backdoor Attacks}
\label{app_attack}

\myparatight{a)~VILLAIN attack~\cite{bai2023villain}}%
The VILLAIN attack is a targeted assault in which the attacker uses a label inference module to pinpoint samples with specific target labels, subsequently embedding a covert additive trigger into them.

\myparatight{b)~He attack~\cite{he2023backdoor}}%
In this attack, the attacker leverages a small collection of labeled target-class data acquired offline and substitutes their local embeddings with a trigger vector during training, enabling control over the final model.

\myparatight{c)~Fu attack~\cite{fu2022label}}%
Fu is a label inference attack aimed at deducing the labels of samples. Once the samples are identified as belonging to the target class, a conventional replacement trigger [1, -1, 1, -1, 1] is applied to create poisoned embeddings.

\myparatight{d)~BadVFL attack~\cite{naseri2024badvfl}}%
BadVFL also starts with a label inference phase, where an auxiliary dataset is used to estimate training labels. Once the source and target classes are identified, a saliency map directs the insertion of a trigger into the features of the source class, generating poisoned embeddings that are moved closer to the target class in the embedding space.

\myparatight{e)~Adaptive attack}%
In the worst-case scenario, where the attacker has full knowledge of the system, specifically our \alg, we devise an adaptive attack strategy targeting \alg. This strategy dynamically adjusts the attack trigger magnitude, \(\lambda\), and uses a binary search approach to iteratively find the maximum \(\lambda\) value. This ensures the attack can influence model updates effectively while evading detection. By doing so, the attacker can carry out the attack without being identified by the defense mechanism, progressively enhancing its impact. This illustrates the strategy's flexibility and adaptability in bypassing defenses. The adaptive attack for our \alg is outlined in Algorithm~\ref{alg:adaptive_attack} in Appendix.

\section{Details of Compared Methods}
\label{app_method}

\myparatight{a)~Trimmed-mean (TrMean)~\cite{yin2018byzantine}}%
The server removes some largest values and smallest values from each dimension of the data, then utilizes the remaining embeddings for training.

\myparatight{b)~Multi-Krum~\cite{blanchard2017machine}}%
Upon receiving the embeddings uploaded by clients and generating the aggregated embeddings, the server selects several candidate aggregated embeddings by minimizing the total distance to their corresponding neighboring subsets.

\myparatight{c)~HDBSCAN~\cite{malzer2020hybrid}}%
Once the aggregated embeddings are generated, the server constructs a density-based hierarchy and identifies stable clusters.

\myparatight{d)~Differential privacy (DP)~\cite{abadi2016deep}}%
During the training phase, the server introduces Gaussian noise with variance \(\sigma_{\text{noise}}^2\) to the aggregated embeddings. These noise-altered embeddings are subsequently used to update the top model, with \(\sigma_{\text{noise}}\) controlling the degree of differential privacy.

\myparatight{e)~Model pruning (MP)~\cite{liu2018fine}}
The server trims the highest weights in each layer of the trained top model by setting them to zero. The pruned model is then employed for inference.

\myparatight{f)~Adversarial neuron pruning (ANP)~\cite{wu2021adversarial}}
The server conducts adversarial pruning by perturbing the trained weights of the top model to pinpoint the most sensitive weights. 
The identified weights are zeroed out, and the pruned model is used for inference.

\myparatight{g)~VFLIP~\cite{cho2024vflip}}%
VFLIP is a server-side masked auto-encoder (MAE)-based defense that cleans abnormal embeddings to mitigate backdoor attacks in SL while preserving inference accuracy.

\section{Details of Non-IID Simulation}
\label{app_noniid}

To model feature-level heterogeneity across clients in SL, we incorporate a controllable parameter $\rho \in [0, 1]$ that specifies the extent to which features are shared among clients. Suppose there are $n$ clients and a global feature space of size $z$. Each client $C_i$ is assigned a subset of features denoted by $\hat{\bm{x}}_k^i$, with $|\hat{\bm{x}}_k^i|$ indicating the number of features allocated to that client.

The simulation begins by considering the global feature index set $\{1, 2, \dots, z\}$. Each client’s local feature set is formed by dividing $|\hat{\bm{x}}_k^i|$ features into two distinct parts: a shared portion common to all clients and a private portion unique to each. The shared component includes $\rho \cdot |\hat{\bm{x}}_k^i|$ features, randomly selected from the global feature space and assigned uniformly to every client. The remaining $(1 - \rho) \cdot |\hat{\bm{x}}_k^i|$ features are drawn from the unused portion of the global space and are distributed to clients in a way that prevents any overlap between them.

For each client, the final feature allocation $\hat{\bm{x}}_k^i$ is constructed by combining the shared features with the client-specific private features. This approach allows systematic control over how much the clients’ feature sets overlap. A setting of $\rho = 0$ ensures completely disjoint feature subsets, simulating a highly Non-IID environment. In contrast, $\rho = 1$ corresponds to identical feature sets across all clients, reflecting an IID condition. Values of $\rho$ between 0 and 1 create partially overlapping configurations, allowing fine-grained tuning of feature-level Non-IID behavior in SL experiments.

\begin{figure*}
    \centering
    \includegraphics[width=1.0\linewidth]{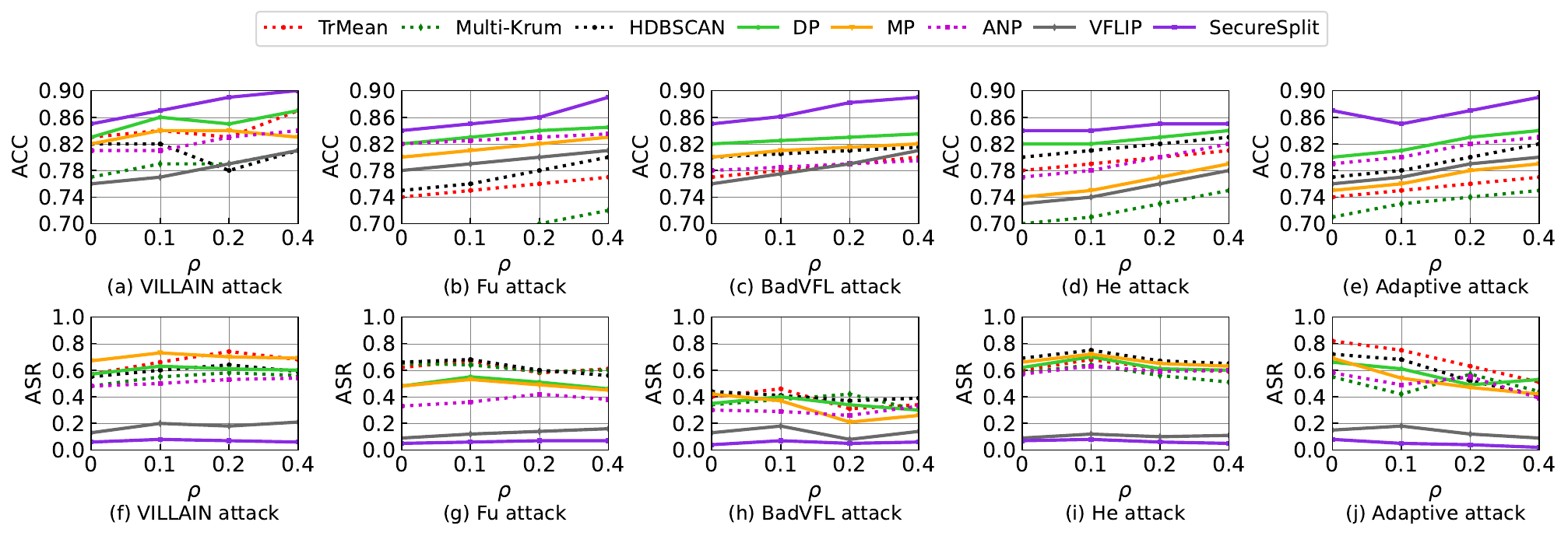} 
    \caption{Impact of degree of Non-IID, where CIFAR-10 dataset is considered.}
    \label{image_overlap}
     \vspace{-0.2cm}
\end{figure*}

\begin{figure*}
    \centering
    \includegraphics[width=1.0\linewidth]{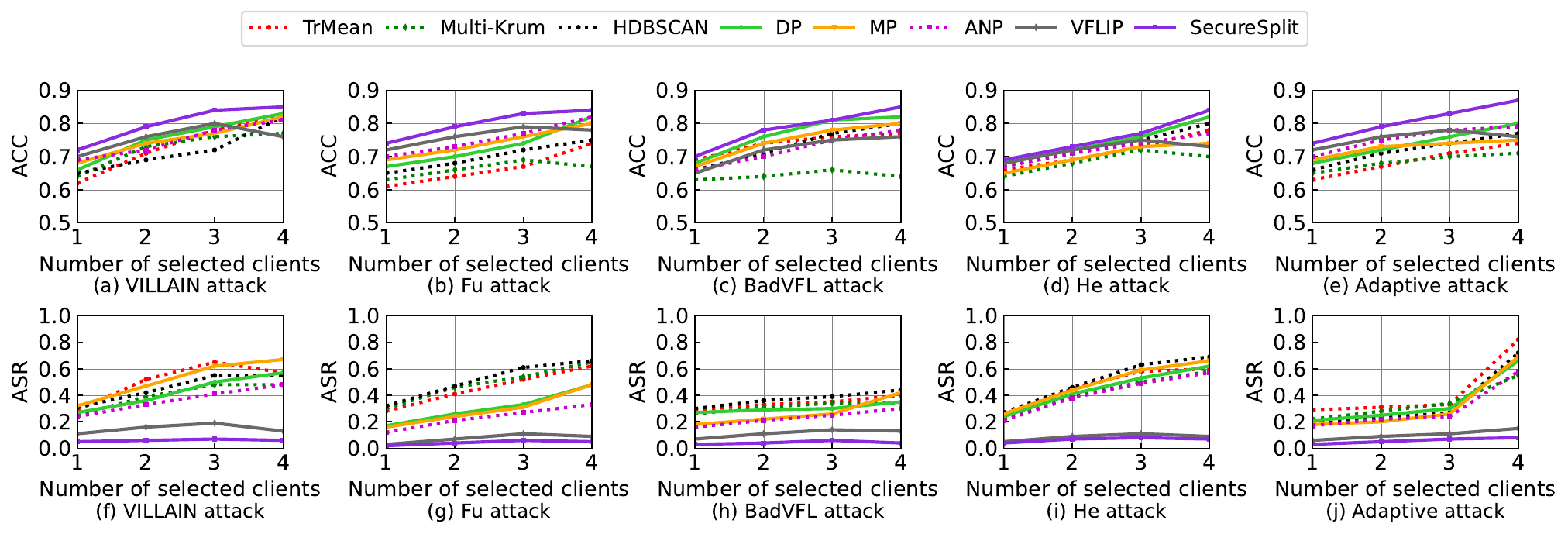} 
    \caption{Impact of the number of selected clients per round, where CIFAR-10 dataset is considered.}
    \label{image_selection}
     \vspace{-0.2cm}
\end{figure*}

\section{More Details of Parameter Settings}
\label{app_more_settings}
The local client model (bottom model) is a 4-layer fully connected network (FCN) for MNIST and CIFAR-10, and a VGG-19 network for ImageNette and CINIC-10. The server-side model (top model) is a 3-layer FCN across all datasets.
All models are trained for 120 rounds. Following the setup in~\cite{he2023backdoor}, the poisoning process begins at the 80th round for all datasets. The learning rate is set to $1 \times 10^{-3}$, with a consistent batch size of 5000 across all datasets. 
In this paper, we set the hyperparameter $\alpha$ to 1.5 by default.

\begin{table*}[t]
\small
\caption{Impact of the poisoning start round. The CIFAR-10 dataset is considered.}
\centering
\begin{tabular}{|c|cc|cc|cc|cc|cc|}
\hline
\multirow{2}{*}{Poisoning start round} & \multicolumn{2}{c|}{VILLAIN attack} & \multicolumn{2}{c|}{Fu attack} & \multicolumn{2}{c|}{BadVFL attack} & \multicolumn{2}{c|}{He attack} & \multicolumn{2}{c|}{Adaptive attack} \\ \cline{2-11} 
                                        & ACC            & ASR          & ACC             & ASR           & ACC           & ASR          & ACC             & ASR           & ACC                & ASR              \\ \hline
20                                      & 0.83          & 0.07         & 0.82           & 0.07          & 0.84         & 0.06         & 0.82           & 0.05          & 0.82              & 0.07             \\
40                                      & 0.85          & 0.07         & 0.83           & 0.09          & 0.83         & 0.07         & 0.82           & 0.06          & 0.85              & 0.04             \\
60                                      & 0.86          & 0.06         & 0.83           & 0.07          & 0.84         & 0.06         & 0.84           & 0.07          & 0.84              & 0.03             \\
80                                      & 0.85          & 0.06         & 0.85           & 0.05          & 0.84         & 0.04         & 0.83           & 0.07          & 0.87              & 0.08             \\
100                                     & 0.86          & 0.03         & 0.86           & 0.03          & 0.86         & 0.03         & 0.85           & 0.02          & 0.86              & 0.07             \\ \hline
\end{tabular}
\label{start_posion}
\end{table*}

\begin{table*}[t]
\centering
\small
\caption{Different variants of \alg. The CIFAR-10 dataset is considered.}
\begin{tabular}{|c|cc|cc|cc|cc|cc|}
\hline
\multirow{2}{*}{Variant} & \multicolumn{2}{c|}{VILLAIN attack} & \multicolumn{2}{c|}{Fu attack} & \multicolumn{2}{c|}{BadVFL attack} & \multicolumn{2}{c|}{He attack} & \multicolumn{2}{c|}{Adaptive attack} \\ \cline{2-11} 
                        & ACC               & ASR              & ACC             & ASR           & ACC               & ASR             & ACC             & ASR           & ACC                & ASR              \\ \hline
Variant I              & 0.78             & 0.61             & 0.80           & 0.45          & 0.80             & 0.60            & 0.81           & 0.62          & 0.81              & 0.67             \\
Variant II             & 0.81             & 0.67             & 0.82           & 0.57          & 0.83             & 0.62            & 0.80           & 0.64          & 0.78              & 0.70             \\
Variant III              & 0.80             & 0.49             & 0.76           & 0.42          & 0.80             & 0.42            & 0.78           & 0.51          & 0.80              & 0.44             \\
Variant IV              & 0.83             & 0.52             & 0.80           & 0.45          & 0.81             & 0.45            & 0.82           & 0.55          & 0.83              & 0.49             \\
Variation V              & 0.82             & 0.04             & 0.80           & 0.01          & 0.79             & 0.04            & 0.79           & 0.03          & 0.82              & 0.04             \\
Variation VI             & 0.83             & 0.19             & 0.83           & 0.17          & 0.84             & 0.22            & 0.82           & 0.21          & 0.85              & 0.22             \\
\cellcolor{greyL} \alg                    & \cellcolor{greyL}0.85             & \cellcolor{greyL}0.06             & \cellcolor{greyL}0.85           & \cellcolor{greyL}0.05          & \cellcolor{greyL}0.85             & \cellcolor{greyL}0.04            &\cellcolor{greyL} 0.83           & \cellcolor{greyL}0.07          & \cellcolor{greyL}0.87              & \cellcolor{greyL}0.08             \\ \hline
\end{tabular}
\label{various_components}
\end{table*}

\begin{table*}[t]
\small
\centering
\caption{The performance of defense methods on various datasets with U-shape SL is evaluated using ACC ($\uparrow$) and ASR ($\downarrow$) metrics, where higher ACC and lower ASR indicate better performance.}
\begin{tabular}{|c|cc|cc|cc|cc|}
\hline
\multirow{2}{*}{Defense} & \multicolumn{2}{c|}{CIFAR-10}    & \multicolumn{2}{c|}{ImageNette}  & \multicolumn{2}{c|}{MNIST}       & \multicolumn{2}{c|}{CINIC-10}    \\ \cline{2-9} 
                         & \multicolumn{1}{c|}{ACC}  & ASR  & \multicolumn{1}{c|}{ACC}  & ASR  & \multicolumn{1}{c|}{ACC}  & ASR  & \multicolumn{1}{c|}{ACC}  & ASR  \\ \hline
No defense               & \multicolumn{1}{c|}{0.67} & 0.94 & \multicolumn{1}{c|}{0.62} & 0.72 & \multicolumn{1}{c|}{0.95} & 0.92 & \multicolumn{1}{c|}{0.65} & 0.87 \\ \hline
TrMean                   & \multicolumn{1}{c|}{0.59} & 0.66 & \multicolumn{1}{c|}{0.57} & 0.49 & \multicolumn{1}{c|}{0.91} & 0.55 & \multicolumn{1}{c|}{0.55} & 0.37 \\ \hline
Multi-Krum               & \multicolumn{1}{c|}{0.57} & 0.58 & \multicolumn{1}{c|}{0.52} & 0.39 & \multicolumn{1}{c|}{0.86} & 0.37 & \multicolumn{1}{c|}{0.54} & 0.29 \\ \hline
HDBSCAN                  & \multicolumn{1}{c|}{0.64} & 0.39 & \multicolumn{1}{c|}{0.56} & 0.52 & \multicolumn{1}{c|}{0.90} & 0.44 & \multicolumn{1}{c|}{0.58} & 0.70 \\ \hline
DP                       & \multicolumn{1}{c|}{0.62} & 0.57 & \multicolumn{1}{c|}{0.60} & 0.44 & \multicolumn{1}{c|}{0.88} & 0.39 & \multicolumn{1}{c|}{0.58} & 0.57 \\ \hline
MP                       & \multicolumn{1}{c|}{0.64} & 0.38 & \multicolumn{1}{c|}{0.59} & 0.62 & \multicolumn{1}{c|}{0.90} & 0.25 & \multicolumn{1}{c|}{0.58} & 0.49 \\ \hline
ANP                      & \multicolumn{1}{c|}{0.59} & 0.44 & \multicolumn{1}{c|}{0.61} & 0.42 & \multicolumn{1}{c|}{0.93} & 0.22 & \multicolumn{1}{c|}{0.60} & 0.39 \\ \hline
VFLIP                    & \multicolumn{1}{c|}{0.56} & 0.38 & \multicolumn{1}{c|}{0.53} & 0.29 & \multicolumn{1}{c|}{0.86} & 0.16 & \multicolumn{1}{c|}{0.52} & 0.18 \\ \hline
SafeSplit                & \multicolumn{1}{c|}{0.66} & 0.32 & \multicolumn{1}{c|}{0.60} & 0.15 & \multicolumn{1}{c|}{0.97} & 0.23 & \multicolumn{1}{c|}{0.60} & 0.27 \\ \hline
\cellcolor{greyL}\alg                     & \multicolumn{1}{c|}{\cellcolor{greyL}0.68} & \cellcolor{greyL}0.04 & \multicolumn{1}{c|}{\cellcolor{greyL}0.64} & \cellcolor{greyL}0.06 & \multicolumn{1}{c|}{\cellcolor{greyL}0.98} & \cellcolor{greyL}0.05 & \multicolumn{1}{c|}{\cellcolor{greyL}0.67} & \cellcolor{greyL}0.08 \\ \hline
\end{tabular}
\label{U_results}
\end{table*}

\section{More Details of U-shaped Split Learning}
\label{app_USL}

\myparatight{a)~U-shaped split learning (U-shaped SL)}%
The U-shaped SL paradigm presented in~\cite{rieger2025safesplit} organizes the model into three sequential components: the head, the backbone, and the tail. The head operates on the client side, where it receives input from the client’s local training data and processes it into an intermediate representation. This intermediate output, often called smashed data, is then forwarded to the server.

At the server, the backbone takes over. It serves as the central computational block of the model, transforming the smashed data into a more refined feature representation. Once this processing is complete, the resulting output is passed along to the tail component.

The tail is also situated on the client side. It takes the processed features from the server, applies the final layers of the model to produce predictions, and calculates the loss if labels are available. The tail then begins the backward pass by sending gradients to the backbone on the server, enabling parameter updates throughout the model.

\myparatight{b)~Adapt the SL method to U-shaped SL architectures}%
To reduce the impact of poisoning attacks, we incorporate defense mechanisms at the stage where the head component communicates with the backbone. This phase involves the transmission of smashed data from the client to the server, making it a critical point for potential adversarial manipulation. To safeguard this process, we apply robust sample filtering techniques that can identify and suppress anomalous or harmful inputs before they influence the server-side computations. Examples of such techniques include TrMean, HDBSCAN, or our proposed \alg.

\begin{figure*}[t]
    \centering
    \begin{minipage}[b]{0.24\linewidth}
        \includegraphics[width=\linewidth]{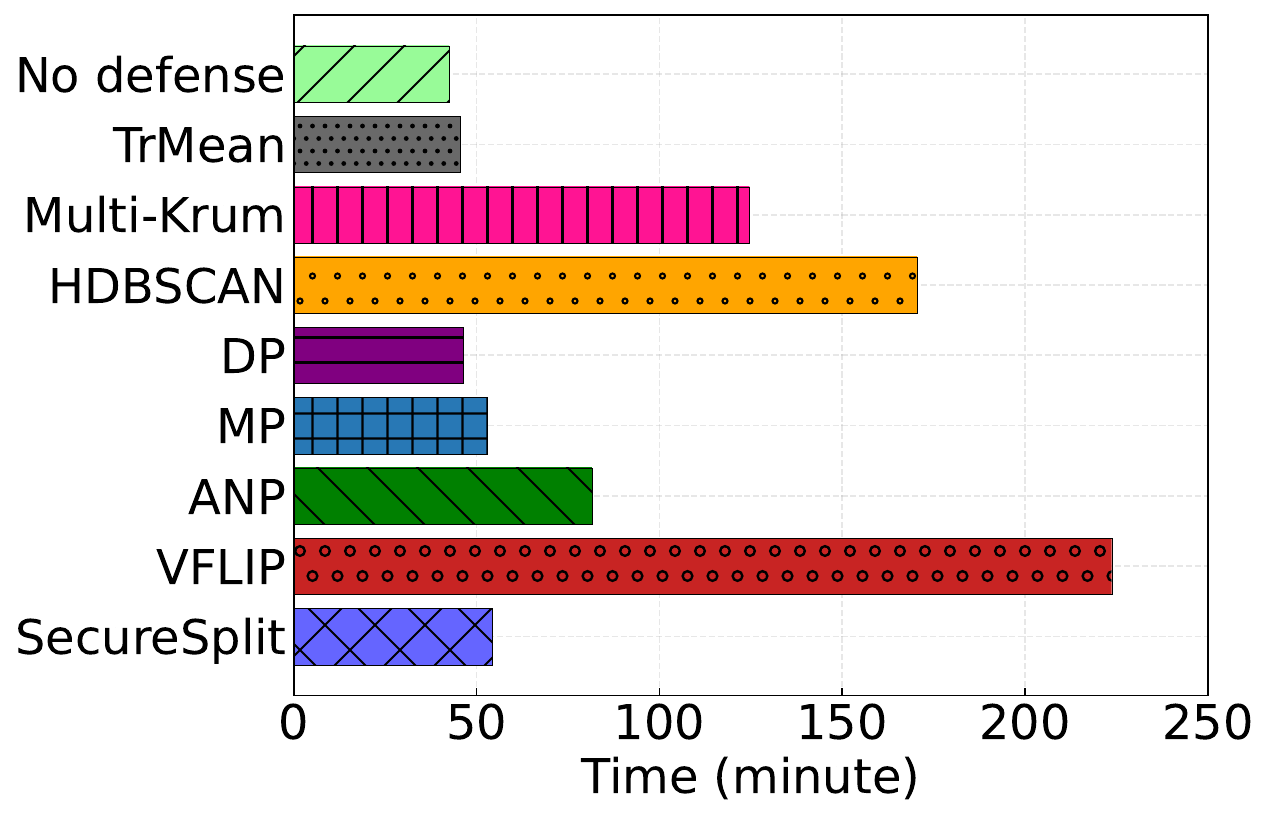}
        \caption{Computation costs of different methods on CIFAR-10 dataset.}
        \label{time_cifar10}
    \end{minipage}
    \hfill
    \begin{minipage}[b]{0.24\linewidth}
        \includegraphics[width=\linewidth]{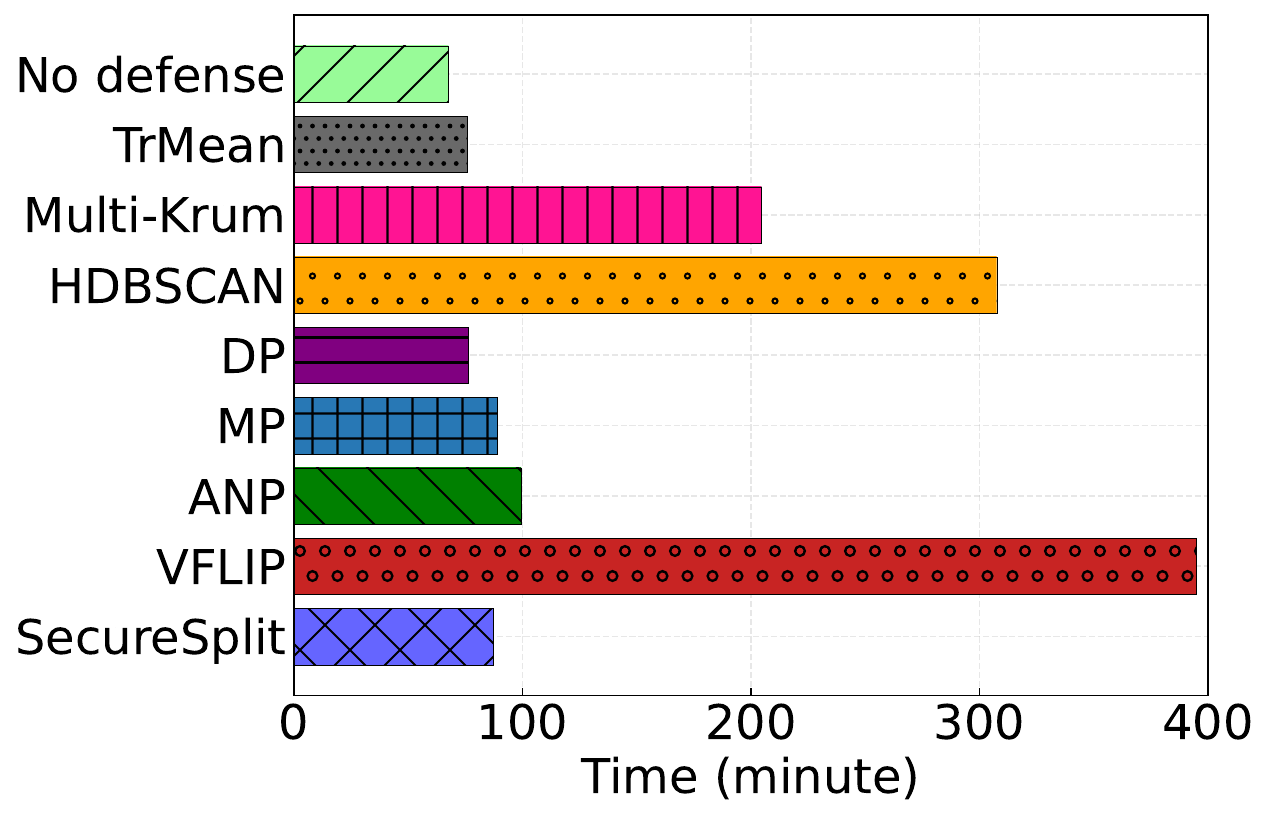}
        \caption{Computation costs of different methods on ImageNette dataset.}
        \label{time_imagenette}
    \end{minipage}
    \hfill
    \begin{minipage}[b]{0.24\linewidth}
        \includegraphics[width=\linewidth]{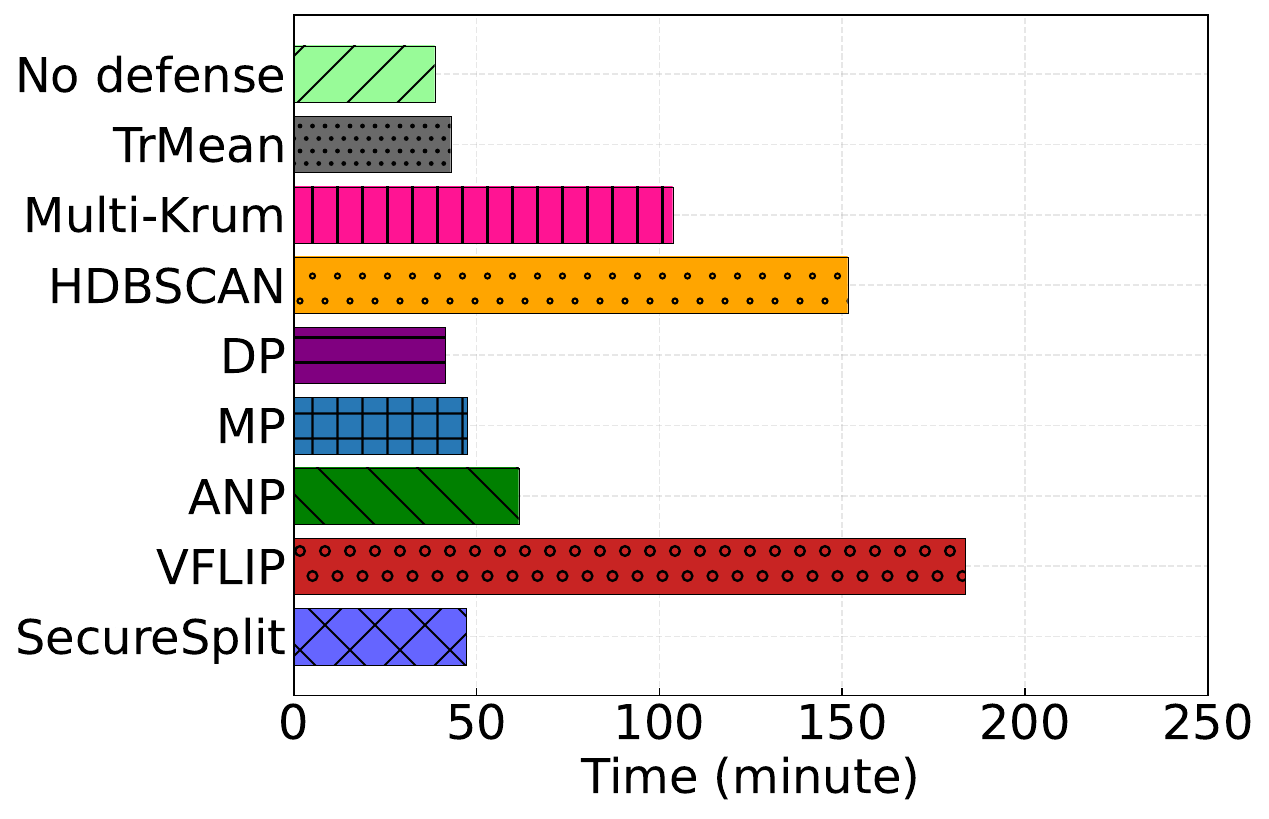}
        \caption{Computation costs of different methods on MNIST dataset.}
        \label{time_mnist}
    \end{minipage}
    \hfill
    \begin{minipage}[b]{0.24\linewidth}
        \includegraphics[width=\linewidth]{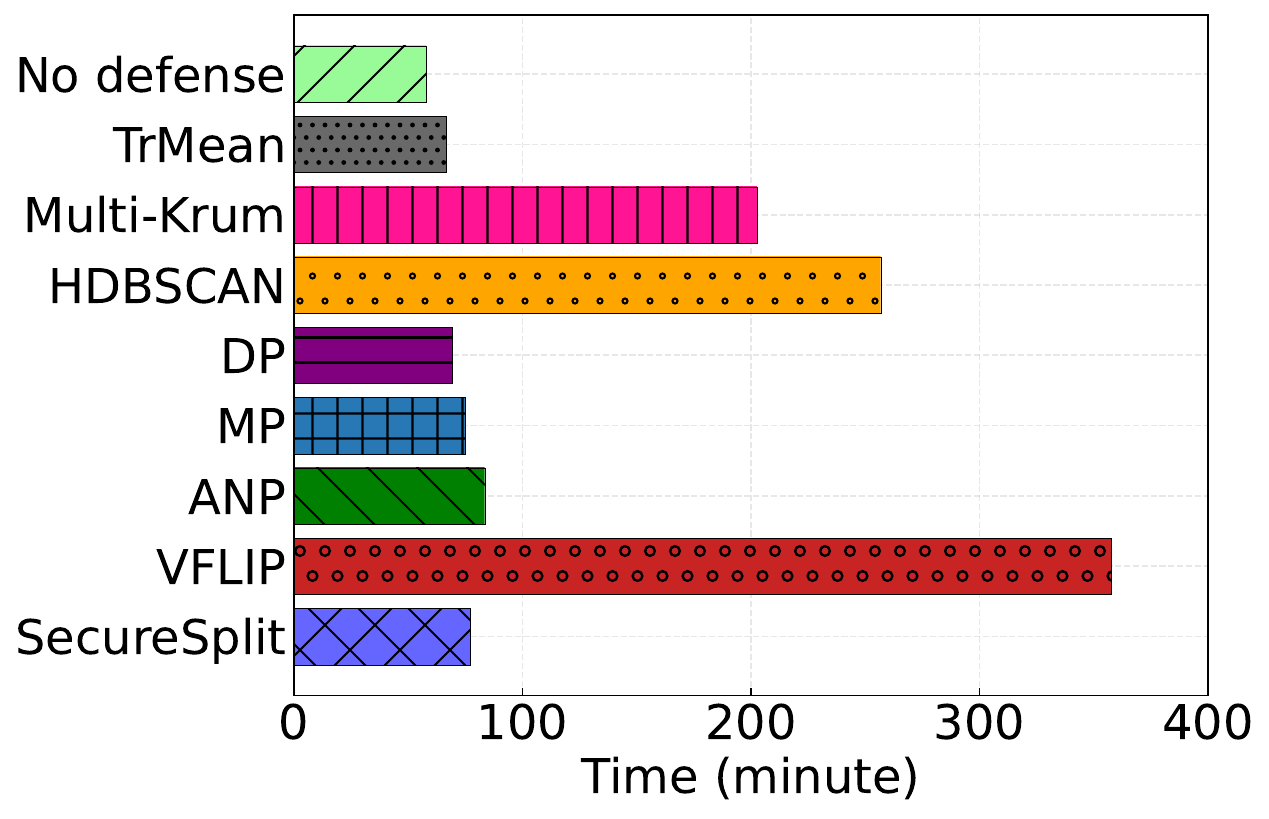}
        \caption{Computation costs of different methods on CINIC-10 dataset.}
        \label{time_cinic10}
    \end{minipage}
\end{figure*}

\end{document}